\documentclass[a4paper,fleqn]{cas-dc}

\usepackage[numbers]{natbib}

\usepackage{graphicx}		
\usepackage{caption}
\usepackage{subcaption}
\usepackage{xcolor}
\usepackage{amsmath}
\usepackage{wasysym}     
\usepackage{float}		
\usepackage{siunitx}		
\usepackage[export]{adjustbox}	
\usepackage{afterpage}    
\usepackage{tabularx}	

\def\tsc#1{\csdef{#1}{\textsc{\lowercase{#1}}\xspace}}
\tsc{WGM}
\tsc{QE}
\tsc{EP}
\tsc{PMS}
\tsc{BEC}
\tsc{DE}

\begin{document}
\let\WriteBookmarks\relax
\def\floatpagepagefraction{1}
\def\textpagefraction{.001}
\shorttitle{\textit{D. Gorkov et~al. / Nuclear Instruments and Methods in Physics Research A XXX (2025) XXX}}
\shortauthors{D. Gorkov et~al.}

\title [mode = title]{KOMPASS: the new cold neutron triple-axis-spectrometer specialized for polarization analysis}                      
%

\author[1,2]{D. Gorkov}[
                        orcid=0000-0002-5983-2771
                        ]
\cormark[1]
\ead{dmitry.gorkov@frm2.tum.de}

\affiliation[1]{organization={II. Physikalisches Institut, Universit{\"a}t zu K{\"o}ln},
                addressline={Z{\"u}lpicher Str.77},                 
                postcode={50937},
                postcodesep={}, 
                city={K{\"o}ln},            
                country={Germany}}
       
\affiliation[2]{organization={Heinz Maier-Leibnitz Zentrum (MLZ), Technische Universit{\"a}t M{\"u}nchen},
                addressline={Lichtenberg Str. 1}, 
                postcode={85748}, 
                postcodesep={}, 
                city={Garching bei M{\"u}nchen},
                country={Germany}}

\affiliation[3]{organization={Physik-Department E21, Technische Universität München},
                addressline={James-Franck-Str. 1}, 
                postcode={85748}, 
                postcodesep={}, 
                city={Garching bei M{\"u}nchen},
                country={Germany}}

\affiliation[4]{organization={Max Planck Institute for Chemical Physics of Solids},
	addressline={N\"othnitzer Str. 40}, 
	postcode={01187}, 
	postcodesep={}, 
	city={Dresden},
	country={Germany}}

\author[2]{M. M{\"u}ller}
\author[2]{G. Waldherr}
\author[1,2]{A. Gr{\"u}nwald}[
                        orcid=0000-0002-9399-5793
                        ]
\author[1]{J. Stein}
[orcid=0000-0001-7136-5419]
\author[3]{S. Giemsa}
\author[1,4]{A. C. Komarek}
[orcid=0000-0002-0084-4698]
\author[3]{P. B{\"o}ni}[
                        orcid=0000-0001-6469-5428
                        ]

\ead{Peter.Boeni@tum.de}
\author[1]{M. Braden}[
                        orcid=0000-0002-9284-6585
                        ]
\ead{braden@ph2.uni-koeln.de}

\cortext[cor1]{Corresponding author}

\begin{abstract}
KOMPASS is a polarized triple-axis cold neutron spectrometer recently installed at the FRM II neutron source. The instrument is designed to operate exclusively with polarized neutrons and is specialized in longitudinal polarization analysis using Helmholtz coils and spherical zero-field neutron polarimetry using a {\sl Cryopad} device.
The advanced guide system polarizes and focuses flexibly in the scattering plane. A first, fixed parabolic focusing part contains a series of three polarizing supermirror V-cavities that produce a highly polarized beam. By exchanging straight and parabolic front-end guide sections, the resolution of the instrument can be optimized to meet experimental requirements.
Large, double-focusing monochromator and analyzer units with pyrolytic graphite crystals enable efficient and adaptive energy selection, and a compact supermirror cavity analyzes the final neutron polarization. Alternatively, or in combination with this cavity, a Heusler polarization analyzer can be used. KOMPASS offers full- and half-polarized configurations with or without secondary energy analysis and provides a wide range of polarization options.
Therefore, KOMPASS is well suited for various studies of static and dynamic magnetic correlations with energy transfers on the neutron energy loss side up to $\approx$12\,meV.

\end{abstract}




\maketitle

\section{Introduction}

Inelastic neutron scattering has proven to be a unique technique to explore excitations in condensed matter systems. Moreover, using either triple-axis- (TAS) or time-of-flight (TOF) spectrometers, very high energy resolution can be achieved. 
In view of the enormous coverage offered by recent TOF instruments one may wonder about
the perspectives of the TAS technique to study a single point of the scattering function 
in the space of scattering vector, $\textbf{Q}$, and energy, $E$, \citep{Shirane2002book}. However, thanks to the continuous development of focusing techniques a TAS yields a more intense signal and a much higher signal-to-noise ratio at the single $(\textbf{Q},E)$ point under study. Therefore, the TAS and TOF techniques are complementary with individual strengths depending on the goal of an experiment aiming either at a small region in $(\textbf{Q},E)$ space and/or a varying external parameter or at a global understanding of the total scattering function.

Exploiting the neutron polarization possesses enormous potential for the investigation of magnetic properties and the distinction between coherent and incoherent scattering. Therefore, first implementations of polarization analysis in neutron scattering were already realized in the 1950ies \citep{TapanChatterji2005book}.
Neutron polarization analysis can not only distinguish between nuclear and magnetic correlations, it can also determine the anisotropy of magnetic correlations in spin space and it can directly 
prove a spiral character of the latter rendering it a most valuable tool
for crystallographic (elastic) as well as for inelastic studies \citep{TapanChatterji2005book}.
With the application of neutron polarization analysis, the superposition of nuclear and magnetic contributions changes from an incoherent to a coherent one thus boosting small magnetic signals. Instead of a single value of the scattering function at $(\textbf{Q},E)$, full 3D polarization analysis yields 36 intensity values, because the scattering function becomes a 3$\times$3 matrix with respect to the initial and final neutron polarization axis, and because each component can be measured with polarization values up-up, up-down, down-up and down-down \citep{TapanChatterji2005book}. \footnote{Note. however, that these 36 informations are not independent.}
However, the existing instrumentation for neutron polarization analysis usually causes a tremendous loss of signal strength in particular for inelastic studies, which represents a true bottle neck for today's magnetism research.

Focusing concepts were developed to enhance the TAS signal in particular for 
measurements with small sample volumes.
In particular, the utilization of a large double-focusing monochromator \citep{Riste1970BentMono} in combination with a horizontally focusing analyzer initially proposed by \citep{Scherm1977variableAna} turned out to be most efficient to enhance the counting rates while even improving the energy resolution.
It was also realized that horizontal focusing at the monochromator should be 
combined with a diaphragm before the monochromator acting as a virtual source, which is, however, difficult
to realize for an existing instrument
\cite{Buehrer1994TASwithVS,Pintschovius1994tas_focusing_mono,Habicht2012optimizationVS}. 
Nowadays, the virtual-source geometry has been implemented in various cold TASs (e.g. PANDA \citep{Schneidewind2007panda}, ThALES \citep{Boehm2015thales2}, IN12 \citep{Schmalzl2016in12} and FLEXX \citep{Skoulatos2011flexx}) and thermal TASs (e.g. PUMA \citep{Sobolev2015puma}, IN20 \citep{Kulda2002in20} and EIGER \citep{stuhr2017eiger}).

Elliptically or parabolically shaped focusing guide sections in the primary spectrometer of a TAS can yield further intensity gain in particular when combined with the virtual-source focusing-monochromator concept.
According to Liouville’s theorem, the product $B = A\cdot\psi$ with $A$ designating the beam area and $\psi$ the solid angle, is conserved  \citep{Boeni2008newconcepts}, therefore these focusing concepts increase the neutron flux at the expense of an enhanced divergence.
Transmitting the required higher divergence by the supermirror (SM) coated guides requires high-performance coatings. 
Thanks to substantial progress in thin-film technologies, one can now produce non-magnetic SM coatings with $m$-values up to $m$ = 8 times the critical angle of nickel  \citep{Schanzer2016SMwithM8}. 
In some recently designed cold TAS instruments,  elliptically tapered guide sections were incorporated in the primary spectrometer \citep{Boehm2008thales,Schmalzl2016in12,Skoulatos2011flexx}, whereas parabolic horizontal focusing was found to be the optimal solution at the KOMPASS instrument \citep{Komarek2011parabolic}.

SM technology can also be applied to study small crystals with dimensions well below the size
of the individual crystals comprising the highly oriented pyrolytic graphite (HOPG) monochromator.  
A compact 
parabolically or elliptically tapered focusing guide with a length of the order of 500\,mm can be installed after the monochromator in front of a small sample. Several built setups \citep{Goncharenko1997focusing,Mirebeau1998focusing,Boeni2014phasespacedensity,Utschick2016focusing,Hils2004focusing,Adams2014focusing,Brandl2015focusing} demonstrated the efficiency of the proposed idea in particular for high-pressure experiments on samples of only a few mm$^3$ volume. Moreover, the  focusing guide also acts as  diaphragms along the flight path thus reducing the background. As a result, the signal-to-noise ratio can be enhanced by more than an order of magnitude. However, this approach leads to an inhomogeneous divergence distribution at the sample position that can lead to a splitting of Bragg peaks (\citep{Boeni2014phasespacedensity}, \citep{Adams2014focusing}, \citep{Brandl2015focusing}). In addition, as emphasized in \citep{Adams2014focusing}, the installation of the focusing guide requires a very careful, reliable and time-consuming alignment. More efficient approaches for beam focusing and phase space selection can be realized using nested mirror optics \cite{Zimmer2019nested}, \cite{Herb2022nested}.


In addition to the enormous progress made in focusing optics, a new polarized TAS can profit from
improved neutron polarizing devices. For instrumentation with cold neutrons, polarizing SM-based optics adapted for large divergences and beam cross-sections, such as S-benders \citep{Skoulatos2011flexx} or multichannel V-cavities \citep{Schmalzl2016in12} seem to be good solutions when combined with large doubly focusing monochromators.

Implementing polarization analysis on a TAS instrument follows the spirit of the TAS technique focusing on a single $({\bf Q},E)$ point and can be realized with high polarization.
Although there are numerous magnetic problems for cold- and thermal-neutron instruments,
the ongoing interest in quantum materials may enhance the demand for experiments with low energy transfers further, which cannot be performed with today's resonant inelastic X-ray scattering (RIXS) instrumentation \citep{AmentRIXS2011}.
Therefore, a dedicated polarized cold-neutron TAS  with the name ``KOMPASS'' was constructed at the FRM II. It already delivered first results before the reactor operation was interrupted \cite{Biesenkamp2021single,Bruning2021cs3fe2br9,Bertin2024interplay,Kunkemoeller2019_SrRuO3}.  
A number of cold-neutron TASs offering optional polarization analysis already operate at the FRM II, ILL, PSI, NIST, ORNL, and ANSTO neutron sources. 
The acronym "KOMPASS" results from the German expression ``\textbf{K{\"O}}ln - \textbf{M}{\"u}nchen für \textbf{P}olarisation \textbf{A}nalyse \textbf{S}pezialisiertes \textbf{S}pektrometer''. It was realized by two groups located at Cologne and Munich with continuous support by the Bundesministerium f{\"u}r Bildung und Forschung.

After defining the general layout of the instrument \citep{Komarek2011parabolic},
the components of KOMPASS were designed and built implementing novel focusing 
and polarization techniques. Here, we present the main characteristics and performances rendering KOMPASS suitable for various experiments on magnetic materials.

\section{Scientific case}

The potential of neutron polarization analysis stems from its inherent ability to distinguish between magnetic and nuclear scattering free of background. 
Therefore, the primary scientific case for KOMPASS lies in the field of complex magnetic materials. 
In the following we provide a non-exhaustive list of scientific problems to be studied with KOMPASS:

- Complex magnetic order is at the core of many problems of topical interest: e.g. multiferroics and magneto-electric materials, magnetic metals, superconductors and frustrated materials \cite{Castelnovo2008SpinIce,Morris2009,Ge2015,Stein2017,Stein2021arrenius}. A complex magnetic structure such as a multicomponent helix  cannot be solved without polarization analysis \cite{TapanChatterji2005book} and in many cases only spherical polarimetry can distinguish between models.
	
- Small magnetic moments are difficult to study, because intensities scale with the square of the moments. Polarization analysis with energy selection enables observation of weak magnetic Bragg reflections or diffuse scattering \cite{Stein2017,Stein2021arrenius,Boeni2002,Schweika2002,Fawcett1988}. For example, small orbital moments were initially reported in high-Tc-cuprates and more recently in an iridate \cite{Mangin2017}. However, these findings remain controversial \cite{Croft2017}. Also, high-pressure experiments with small sample volumes will profit from the high signal-to-noise ratio that is provided by a cold polarized TAS.
	
- Magnetic frustration can lead to new states of matter such as the long-searched quantum spin liquids. They exhibit new types of excitations \cite{broholm2020,kitaev2006,Castelnovo2008SpinIce,trebst2022}. Besides the possibility to study fundamental problems such as fractionalization, the peculiar character of the excitations may contribute to developments in quantum computing \citep{kitaev2006}. 
	
- Topological concepts change our general approach to condensed matter phenomena, but only recently it was discovered that there is a direct impact on the spin dynamics \citep{Itoh2016, Jenni2019}. Weyl points can renormalize a ferromagnetic magnon dispersion and eventually even invert the magnon handedness \cite{Jenni2022}. The concept of topological magnons was confirmed very recently \cite{McClarty2022,Itoh2016,Nagaosa2010,Zhu2021}. 
	
- Magnetoelectrics and multiferroics attract interest due to their potential in information storage and processing \cite{spaldin2019,Finger2010electric,Biesenkamp2021chiral,Biesenkamp2021single}. Dynamic magnetoelectric coupling results in non-reciprocal properties \cite{Bordacs2012} and in hybrid magnon-phonon modes, labeled electromagnons (see e.g. \cite{Senff2007TbMnO3,Senff2008TbMnO3_magnon_spectrum})). However, a general understanding is missing.
	
- The observation of skyrmions and skyrmion lattices had a tremendous impact after their discovery in MnSi \cite{Muehlbauer2009skyrmion} opening a possible path for future
technologies. Besides the static arrangement studied by neutron diffraction, examining the dynamics of skyrmions is a key for future applications.
	
- Quantum phase transitions and quantum criticality \cite{Sachdev1999_QPhaseTransitions} describe an intriguing quantum-mechanical problem, where energy and temperature merge to a single scale that is possibly related to unconventional superconductivity \cite{Scalapino2012_UnconvSCs}.
	
- Strong spin-orbit coupling: For $4d$ and $5d$ materials spin-orbit coupling can be sufficiently strong to determine the magnetic state \cite{Pesin2010,Witzak2014,Banerjee2017neutron}. Even for an intermediate strength the magnetic interaction will become anisotropic requiring studies with polarized neutrons \cite{Steffens2013optDopedBaFeCoAs,Steffens2019_Sr2RuO4,Qureshi2012_anis_BFA}. The results may be interpreted in terms of the Kitaev model or Majorana fermions \cite{trebst2022,braden2024}. 
	
- Unconventional and topological superconductivity form a fascinating field tightly connected to magnetism. Magnetic excitations only occurring in the superconducting state, so-called spin-resonance modes, are interpreted as fingerprint of a magnetic pairing mechanism \cite{Scalapino2012_UnconvSCs,Steffens2013optDopedBaFeCoAs,Wasser2017anis_res_mode,Wasser2019spin_res_mode}. 
	
- Time resolved neutron studies can be performed with stroboscopic techniques yielding a time resolution of $\sim$50 micro seconds \cite{Stein2021arrenius,Biesenkamp2021chiral}. So far, electric fields were used to control the magnetic domains most efficiently in a multiferroic. Further efforts are needed to rapidly manipulate the samples by magnetic fields or currents. 
	
- Itinerant magnetism  typically results in small ordered moments and correspondingly small cross sections for the excitations, which furthermore can be anisotropic, as in  BaFe$_2$As$_2$ and Sr$_2$RuO$_4$ \citep{Qureshi2012_anis_BFA, Steffens2019_Sr2RuO4}. The presence of longitudinal modes are little explored so far \citep{moriya1979recent, boeni1991longFluctNi, boeni1995EuS}.
	
- Nematic and cholesteric phases: In FeAs based superconductors and in some ruthenates the role of nematic phases is well established \cite{Bohmer2022} while their relevance remains unclear in many other systems. Polarized INS experiments under uniaxial pressure are most promising to resolve these issues. However, they are technically challenging.

- Excitations in thin films and multilayers are difficult to observe due to the small sample volume \cite{Grunwald2010,Grunwald2010b}. However, combining polarization analysis and background reduction with the analyzer as well as focusing techniques will facilitate inelastic experiments. 
	
- Altermagnetism was only recently realized as a class of magnetic materials, which combine aspects of ferromagnetic and antiferromagnetic order opening the path to spintronics applications \cite{Smejkal.2022,Smejkal.2022b}. Polarized neutron experiments can clarify the magnetic structure and most importantly may confirm the predicted splitting of chiral magnons \cite{Smejikal.2023}.

\section{General layout of KOMPASS}

\begin{figure}[h]
	\centering
	\includegraphics[width=.9\columnwidth]{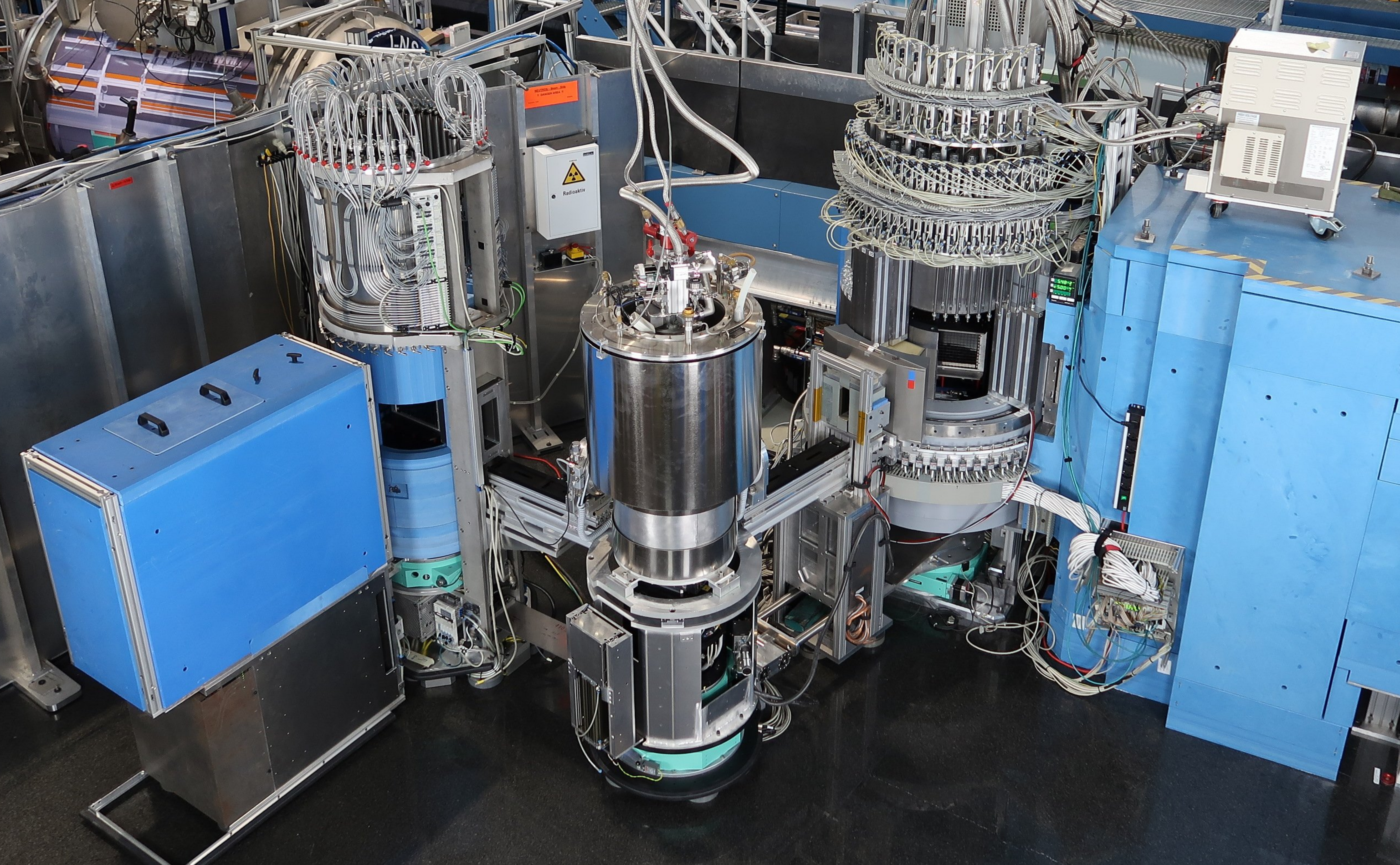}
	\caption{Overview of the KOMPASS spectrometer. On the sample table, a closed-cycle refrigerator is mounted inside the {\sl Cryopad}. The instrument area is shielded by a fence of shielding walls (gray color). The borated polyethylene shielding of the beamline appears in blue.}
	\label{fig:KOMPASS TAS}
\end{figure}

The instrument KOMPASS (Fig. \ref{fig:KOMPASS TAS}) is located at the FRM II in the neutron guide hall West at a distance of 45\,m away from the cold source. The instrument occupies the end position of the curved neutron guide NL1 which has dimensions of 60 $\times$ 120\,mm ($w \times h$). For an improved transport of thermal neutrons, the curved section of the NL1 is subdivided into two channels of equal width, resulting in a critical wavelength $\lambda$\textsubscript{crit} = 1.8\,\AA{} (corresponding to an energy of $E = 25$ \,meV) \citep{zeitelhack2006NL1} in front of the KOMPASS polarizer. 

Special emphasis during the design of KOMPASS was put on a compact layout involving short distances between the monochromator–sample and sample–analyzer of about 1.2\,m only and on the implementation of advanced focusing techniques throughout the instrument. 
The distance between the analyzer and the detector can be varied between 1\,m and 1.3\,m depending on the selected detector unit as discussed in section \ref{sec:Detector_options}. 

For experiments requiring high ${\bf Q}$ resolution, the mono\-chromator–sample and sample–analyzer distances can be enlarged up to 1.67 and 1.3\,m, respectively, and optional collimators can be inserted at three positions along the neutron path. Furthermore, narrow diaphragms can be placed in front of the detector.
%

\begin{figure*}[]
\centering
     \begin{subfigure}[b]{1\textwidth}
		\hspace*{0cm} 
		\includegraphics[width=17.5cm,keepaspectratio]{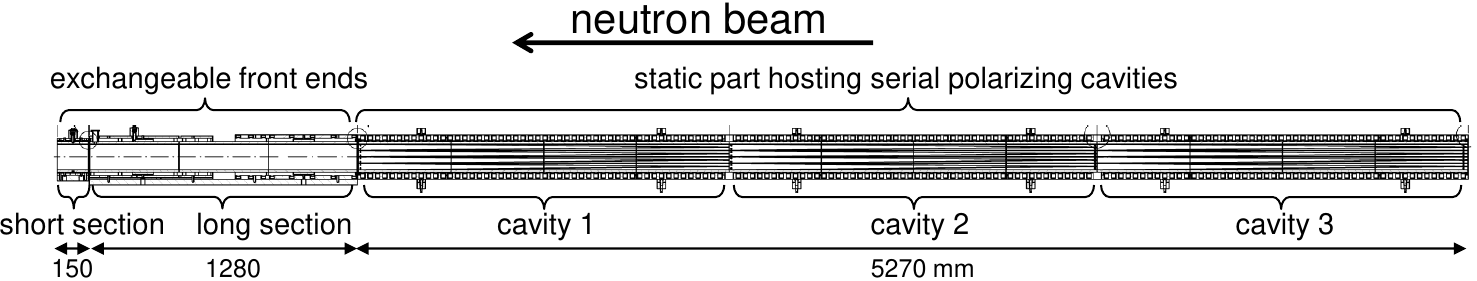}
		\caption{}
		\label{fig:KOMPASS scheme triple V-cavity}
     \end{subfigure}
\hfill		
	\begin{subfigure}[b]{1\textwidth}
		\centering
		\vspace*{0.5cm}
		\includegraphics[width=\textwidth]{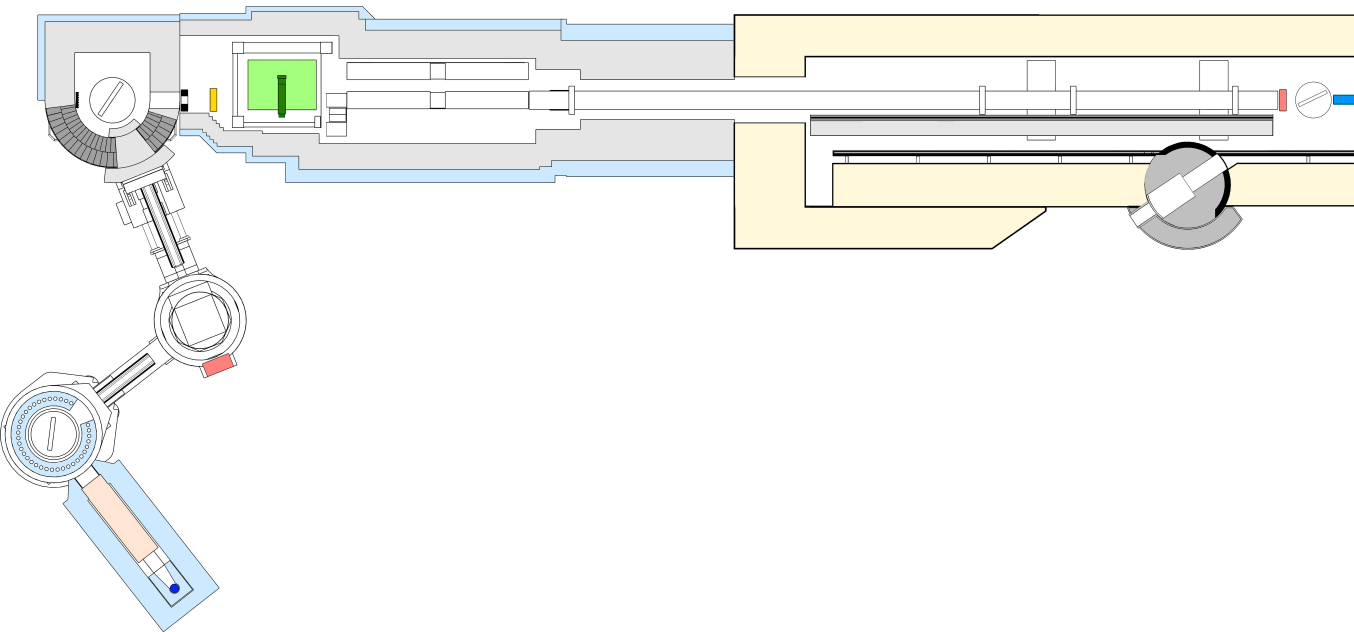}
		\caption{}
		\label{fig:KOMPASS scheme all}
	\end{subfigure}
	\begin{subfigure}[t]{1\textwidth}
		\vspace*{-9.1cm}
		\includegraphics[width=\textwidth,right]{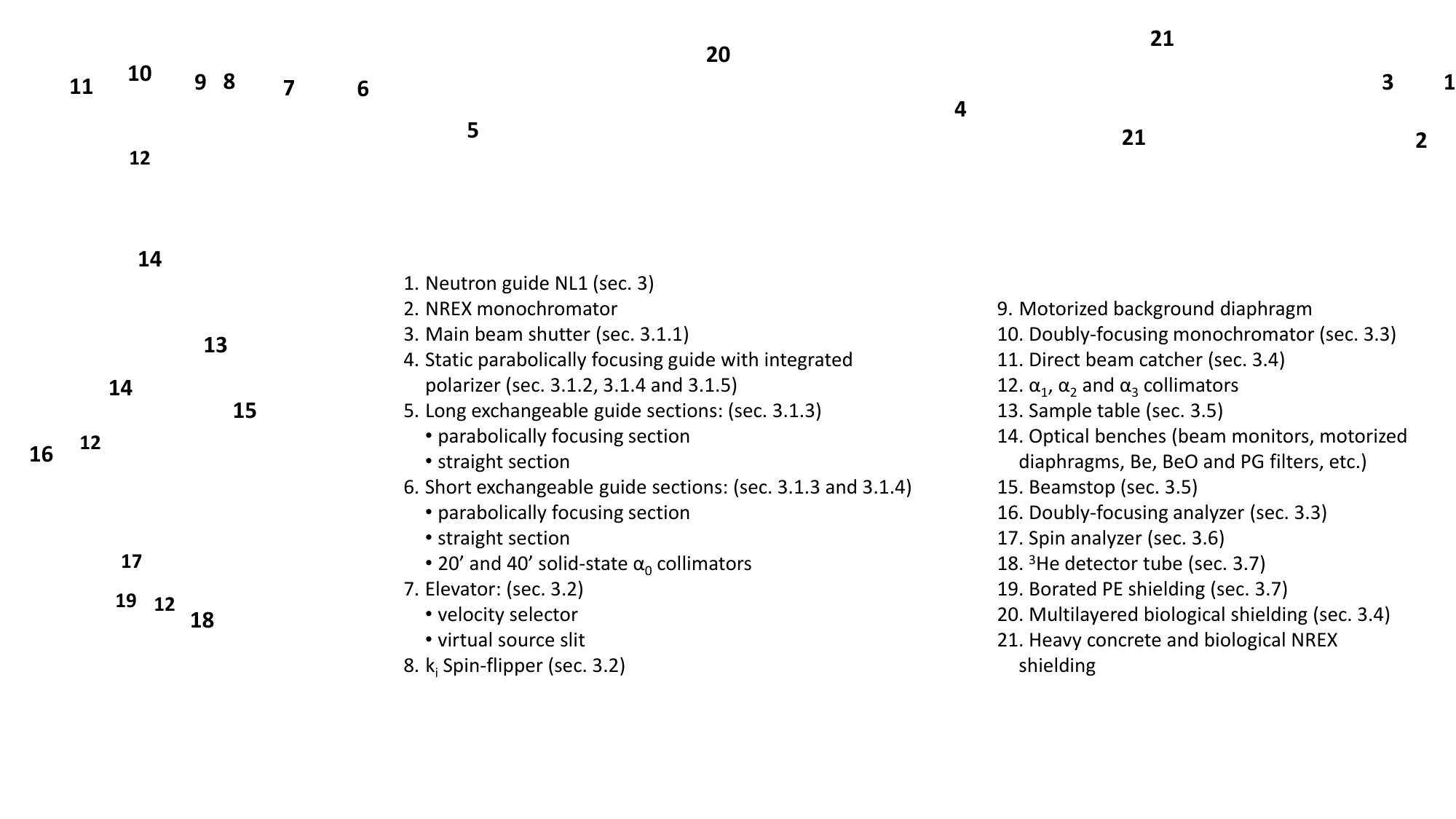}
	\end{subfigure}
%
	\caption{
	(a) Side view of the polarizing guide system showing the fixed and exchangeable sections. (b) Scheme of the KOMPASS spectrometer. The numbers in brackets designate the number of the respective sections where more detailed information about the component is given.}
\end{figure*}

Figures \ref{fig:KOMPASS scheme triple V-cavity} and \ref{fig:KOMPASS scheme all} show details of the polarizing triple-cavity (top) and a layout of the complete KOMPASS spectrometer (bottom). The numbers in brackets provide links to the sections where the designated components are described in detail.

\subsection{Adaptable polarizing neutron guide system} 

Monte-Carlo simulations using the neutron ray-trace simulation package McStas (\citep{Lefmann1999mcstas}, \citep{Nielsen2000monte}, \citep{WILLENDRUP2011S150}) were used to define and optimize the beamline including the given guide system up-stream of the triple cavity. In particular, the polarizing guide system and the exchangeable front ends were optimized based on the simulations with McStas.

The KOMPASS guide system is divided into three sections comprising a total length of 6.7\,m. The first, permanently installed section of 5.27\,m length  
is parabolically focusing in the horizontal dimension containing polarizing V-cavities delivering a polarized beam. Downstream it is followed by a long ($L_1 = 1.28$ m) and a short ($L_1 = 0.15$ m) section, In these sections either parabolically focusing elements or straight guides can be remotely translated into the beam. This way four different combinations for focusing can be realized. However, only three of them provide a reasonable beam definition, as the combination of the long-straight with the short-parabolically focusing section is of no use.

The variable optical configurations allow selecting an optimum illumination of the sample as required by the experiment. A TAS spectrometer positioned at the end position of a parabolically focusing neutron guide in combination with a doubly-focusing monochromator promises a high intensity and and a superior energy resolution when compared with a straight or elliptic neutron guide \citep{Komarek2011parabolic}. The price for this is, of course, a reduced transverse ${\bf Q}$ resolution. Furthermore, the inhomogeneous divergence ot the neutron beam at the sample is reduced when compared with elliptic focusing (\citep{Komarek2011parabolic}). 

In the following subsections, a detailed description of the components comprising the guide system of KOMPASS is given (see Figures \ref{fig:KOMPASS scheme triple V-cavity} and \ref{fig:KOMPASS scheme all}). It starts with the primary neutron-beam shutter and ends with a description of the magnetic guide fields.


\subsubsection{Main beam shutter}

The neutron guide NL1 ends with a short gap of 400\,mm that is required for the monochromator of the up-stream instrument NREX \citep{khaydukov2015nrex}. At the end of the gap, the main neutron beam shutter of KOMPASS is installed. It is made from a 8\,mm thick B\textsubscript{4}C plate enriched with \textsuperscript{10}B up to at least 95\,\% and a 10\,mm thick lead plate behind suppressing the neutron-capture gamma radiation. Immediately after the shutter, the permanently installed polarizing guide begins.

\subsubsection{Static part of the polarizing guide system}

The permanently installed static part of the guide system hosts a series of three polarizing multichannel transmission V-cavities, each being 1.75\,m long. The sides of the cavities are made from borofloat glass, except those of the first V-cavity, which are exposed to the highest neutron flux. Therefore, more radiation-resistant borkron glass N-BK7 was used instead. Each of the three supermirror V-cavity systems consists of a vertical stack of four 30\,mm high V-elements as shown in the side-view of the polarizing guide system (Fig. \ref{fig:KOMPASS scheme triple V-cavity}).

Each of the V-elements hosts a pair of horizontally inclined SMs separated by horizontally installed thin absorbing blades. The SMs consist of silicon wafers that are coated on both sides with a Fe/Si SM with $m = 4.2$. The separating blades of the first cavity are made from silicon wafers that are coated on both sides with a thin absorbing TiB\textsubscript{2}/Ti multilayer (in total corresponding to \SI{0.5}{\micro\m} TiB\textsubscript{2}) subsequently coated with non-magnetic Ni/Ti SM ($m$ = 1.6). 
The absorbing blades of the second and third cavity consist of Desag\,263\,T borosilicate glass that is coated on both sides with non-magnetic SM ($m$ = 1.6).

\begin{figure}
     \centering
     \begin{subfigure}[b]{1\columnwidth}
         \centering
         \includegraphics[height=2.1cm,keepaspectratio]{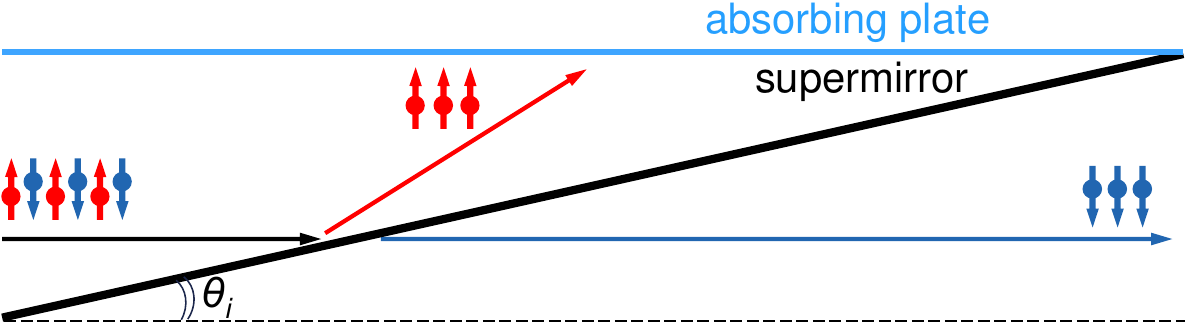}
         \caption{}
         \label{fig:cavity scheme1}
     \end{subfigure}
\hfill		
         \label{fig:cavity scheme2}
%
     \begin{subfigure}[b]{1\columnwidth}
         \centering
     \includegraphics[height=1.9cm,keepaspectratio]{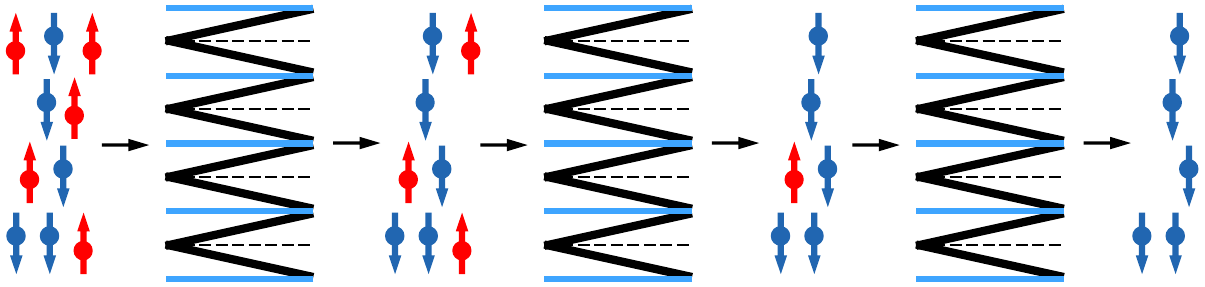}
         \caption{}
         \label{fig:cavity scheme3}
     \end{subfigure}
\hfill		
\begin{subfigure}[b]{1\columnwidth}
        \centering
    \includegraphics[width=8.4cm,keepaspectratio]{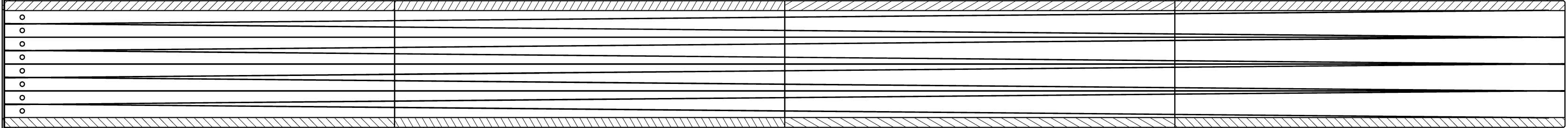}
         \caption{}
         \label{fig:cavity scheme4}
    \end{subfigure}
        \caption{Panel (a) explains the ideal process of polarizing an incident unpolarised neutron beam by means of a Si-wafer that is coated with polarizing Fe/Si supermirror. It is inclined under an angle of $\theta_i$ with respect to the beam direction. (b) Schematic of the polarizing triple V-cavity (side view). The vertical division of the guide body into 4 V-channels reduces the length of the whole polarizer by a factor of four. Note that the vertical scale of the panel is stretched. (c) Technical drawing of one of the V-cavities with a height of 120\,mm. The cavity is assembled from 4 elements yielding the total length of 1750\,mm.}
        \label{fig:cavity scheme}
\end{figure}

The polarizing effect of a V-cavity is based on the spin-dependent reflection of an unpolarized neutron beam from the polarizing SM that is placed under a shallow angle $\theta_i$ with respect to the incident beam (Fig. \ref{fig:cavity scheme}(a), black solid line).

Within a range of angles of incidence that is determined by the $m$ value of the polarizing SM coating, the spin-up component of the beam is reflected from the SM and becomes absorbed by the absorbing blades (blue solid line), or by the sides of the guide. At the same time, the spin-down component is transmitted through the SM-coated silicon wafer, which has a thickness of 0.3\,mm. For doubling the accepted beam cross-section, two inclined SMs are combined into a single V-element. To avoid leaking of neutrons with spin-up polarization, the tips of the V-elements overlap and the ends intrude the sides of the guide or overlap with the absorbing blades.

However, the range of angles of reflection of the polarizing SMs is rather small, i.e. a few degrees only, resulting in small inclination angles $\theta_i$. Therefore, the V-cavities have to be very long. A height of the beam of $h = 120$ mm results in a length of the cavity with one V-channel of 7000 mm. By arranging the V-elements, each of 30 mm height, in four parallel channels, the effective length of the cavity can be reduced to 1750 mm (Fig. \ref{fig:cavity scheme}(b)).

\begin{figure}[h]
	\centering
	\includegraphics[width=.85\columnwidth]{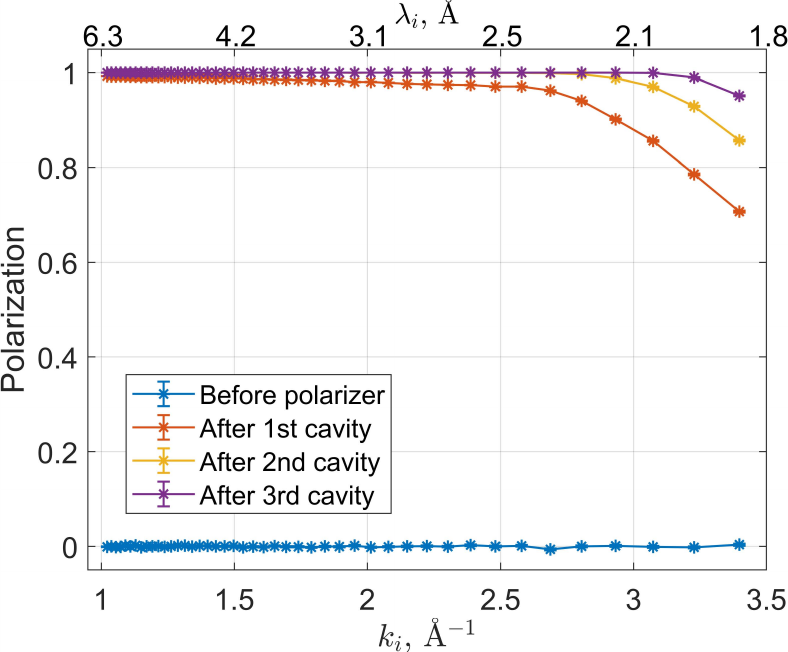}
	\caption{Simulated polarization of the beam subsequently transmitted through one, two, and three rectangular polarizing multichannel V-cavities that are placed after the main beam shutter.}
	\label{fig:Sim_Pol_VVV_Cavity}
\end{figure}

The polarizing efficiency can be further enhanced via serial stacking of several multichannel cavities, which becomes more pronounced with increasing neutron energy. Figure \ref{fig:Sim_Pol_VVV_Cavity} shows the results of a McStas simulation providing the polarization of the originally unpolarized beam after passing one, two, and three 4-channel V-cavities. The simulations reveal a very high degree of polarization $P\simeq 1.0$ across the whole beam area in a range of wavelengths 2 \AA\ $\le \lambda \le$ 6.3 \AA. Moreover, the transmission of the polarizer as given by the number of the polarized neutrons leaving the cavity divided by the number of the incident (unpolarized) neutrons is close to 0.25.

By technical reasons, the maximum length of guide elements is restricted to be 500 mm. As shown in Figure \ref{fig:cavity scheme}(c), four elements are glued together to yield a polarizing cavity with a length of 1750 mm and a height of 120 mm. Note that the sides of the cavities have a parabolic profile in order to provide horizontal focusing.


\subsubsection{Variable part of the guide system}

The variable part of the guide system is composed of four exchangeable guide sections which are mounted in pairs on two successive translation tables. Each pair consists of one focusing and one straight guide. The guide sections of the first and second pair are 1.28\,m and 0.15 m long, respectively. The coatings of the guide sections are given in Table \ref{tab:guide_coating}. They can be translated into the beam allowing the adaption of the energy- and momentum-resolution according to the needs of the experiment. Highest flux combined with a high energy- and low momentum-resolution is obtained with the configuration, where both parabolically focusing long and short sections are translated into the beam. In contrast, for measurements requiring high transverse-momentum resolution, i.e. for investigating excitations with a steep dispersion, the straight neutron guides are selected. To improve the ${\bf Q}$ resolution further, the short guide can be replaced by a 20’ or 40’ Soller-type solid-state collimator.
 
\begin{table*}[width=.765\textwidth]
\caption{Distribution of the $m$-value of the non-depolarizing supermirror coating. With increasing distance from the entrance of the guide, the $m$-value of the sides increases due to the increasing divergence of the neutron beam.}
\label{tab:guide_coating}
\begin{tabular*}{\tblwidth}{@{} |l|c|c|c|c|c|c|c|@{} }
%
\toprule
\multirow{3}{*}
{} & \multicolumn{3}{c|}{Serial polarizing cavities} 
                  & \multicolumn{4}{c|}{Exchangeable front-ends}\\
\cline{2-8}
{} & \multirow{2}{*}{Cavity 1} & \multirow{2}{*}{Cavity 2} & \multirow{2}{*}{Cavity 3}
   & \multicolumn{2}{c|}{Long sections}
                                    & \multicolumn{2}{c|}{Short sections}\\
\cline{5-8}
{} & {}                        & {}                        & {}
   & {Focusing}& {Straight}
                                    & {Focusing}& {Straight}\\
\midrule
Side	       &2.5   &3    &3.5 and 4	&5 and 6 &3.5            &6 &3.5  \\
\hline
Top/Bottom &1.6   &1.6  &1.6         &\multicolumn{2}{c|}{2}  &\multicolumn{2}{c|}{2}   \\
\bottomrule
\end{tabular*}
\end{table*}

For finest ${\bf Q}$ resolution the horizontal divergence of the beam can be further reduced by inserting Gd\textsubscript{2}O\textsubscript{3}-coated-Mylar-foil Soller collimators. The technical details of the collimators are listed in the Table \ref{tab:technical_specification} of the Appendix. They can be placed after the monochromator, in front of the energy analyzer and in front of the detector (see item 12 in Fig. 2(b)).

For reducing the instrument background due to scattering by air, the polarizing cavities and the exchangeable guide sections are equipped with thin aluminum windows and are evacuated down to 10\textsuperscript{-2}\,mbar pressure by means of a scroll pump.

\subsubsection{Optimization of the supermirror coating}

The sides of the guide system are coated with non-depolarizing SM with the index $m$. With increasing distance from the entrance of the guide it increases in the static part progressively from 2.5 to 4 due to the decreasing width $w$ of the parabolic profile according to Liouville's theorem, i.e. $wm \simeq \rm{const}$. It further increases from $m$ = 5 at the beginning of the long exchangeable section to $m$ = 6 at the end of the short focusing section. 

On average, the sophisticated  $m$-value variation in the  sides from 2 to 6 corresponds to a uniform $m$ = 3.5 coating. McStas simulations show that the guide system delivers about 10\% more intensity when compared with an identical system coated uniformly with $m$ = 4 \citep{Komarek2011parabolic}. Moreover, as the number of layers of a SM is proportional to $m$\textsuperscript{4} \citep{Boeni1997sm_devices} and the area for large-$m$ coatings is rather small, the present guide system is more cost-efficient than a system with a uniform coating.

For the sides of the straight long and short sections, the uniform coating with $m$ = 3.5 was found to be optimal for the neutron transport. The top and bottom of the guide are uniformly coated with non-depolarizing SM $m$ = 1.6 for the static parts (including the dividing blades) and with $m$ = 2 for the movable guides. The $m$-value distribution along the guide system is summarized in  Table \ref{tab:guide_coating}.

\subsubsection{Magnetic guide fields}

The cavities and the guide sections are surrounded by magnetic casings that provide a magnetic field for saturating the magnetization of the Fe/Si coatings. Its strength of approximately 70 mT ensures the absence of residual domains in the polarizing SMs resulting in a highly polarized and simultaneously highly intense neutron beam provided by the SM cavities. The very first tests performed at KOMPASS with an opaque \textsuperscript{3}He cell confirmed the high quality of the incident beam with polarization of above $P = 97.4\,\%$ over the whole beam cross-section. $P$ will be further improved by optimizing the guide fields between the exit of the guide system and the sample position.\footnote{Optimizing the guide fields can only be realized after restarting the FRM-II reactor.} We point out, that $P = 97.4\,\%$ corresponds to a flipping ratio $R = 76$ that is defined as the ratio of non-spin-flip to spin-flip intensity measured at the sample position. It is related to the incident polarization by $P= (R-1)/(R+1)$.

The high polarization is secured by guide fields that are installed along the entire beam path on the way to the spin-analyzer. Because the SMs  in the polarizing cavities are inclined with respect to the horizontal plane, the saturating fields are applied horizontally i.e. parallel to the SM surfaces.

To facilitate the handling of the polarization after the triple-cavity, the polarization of the beam is rotated from the horizontal to the vertical direction. This task is accomplished by splitting the magnetic housing of the long guide sections in two parts, separated by a gap of 100 mm. The upstream part of the magnetic housing generates a horizontal field, whereas the following part generates a vertical one. Within the gap, the rotating magnetic field always exceeds 16 mT thus guaranteeing an adiabatic rotation of the polarization due to the adiabaticity parameter being at least 85 for neutrons with $\lambda = 1.8$\,\AA{}, which represents the high-energy limit of the neutron spectrum at KOMPASS. At the entrance of the spin-analyzing cavity of the secondary spectrometer the polarization rotates back to the horizontal direction.

\subsection{Velocity selector and neutron spin-flipper} 

\begin{figure}[htb]
     \centering
     \begin{subfigure}[b]{0.45\columnwidth}
         \centering
         \includegraphics[height=3.5cm,keepaspectratio]{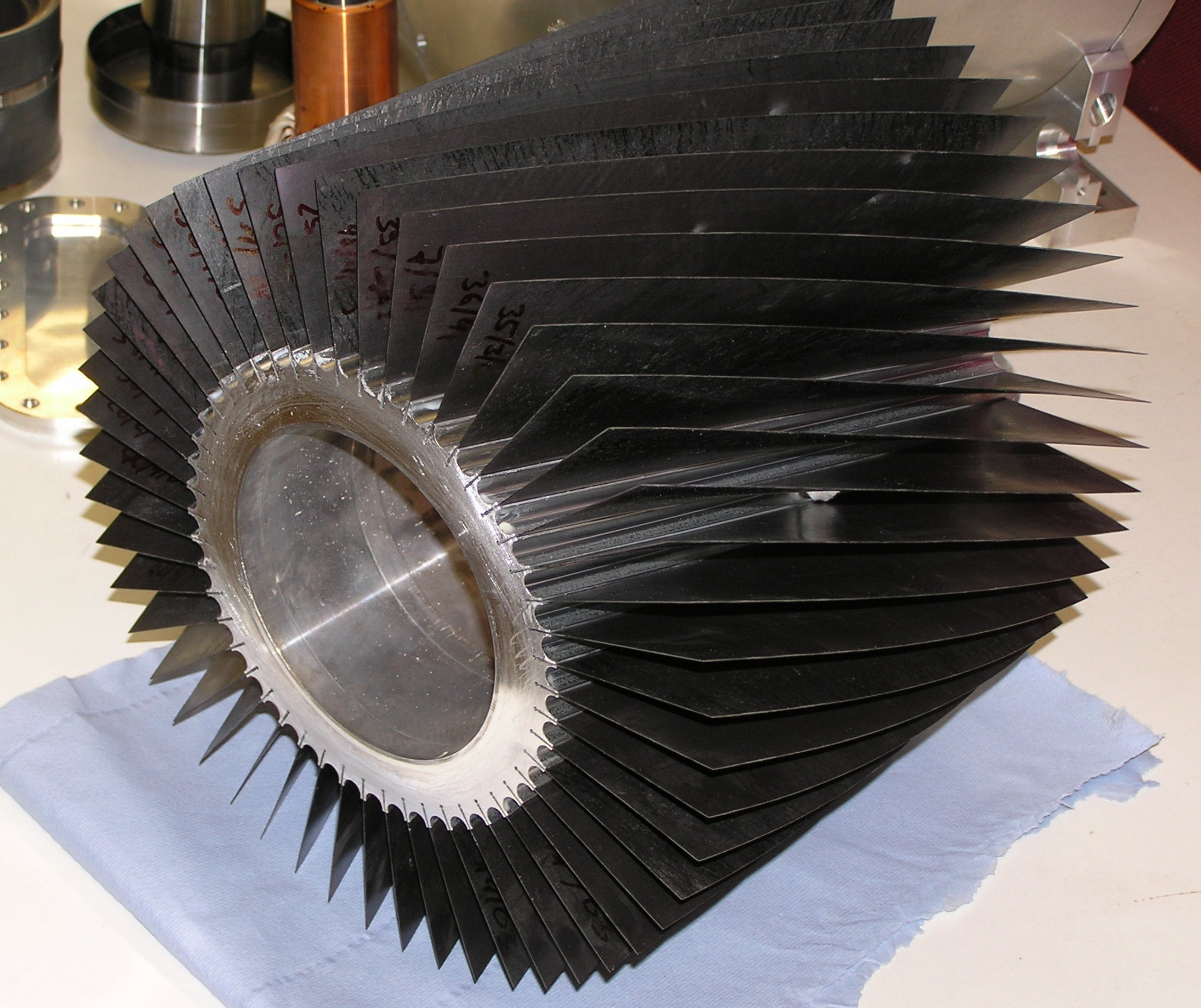}
         \caption{}
         \label{fig:selector1}
     \end{subfigure}
     \begin{subfigure}[b]{0.45\columnwidth}
          \centering
         \includegraphics[height=3.5cm,keepaspectratio]{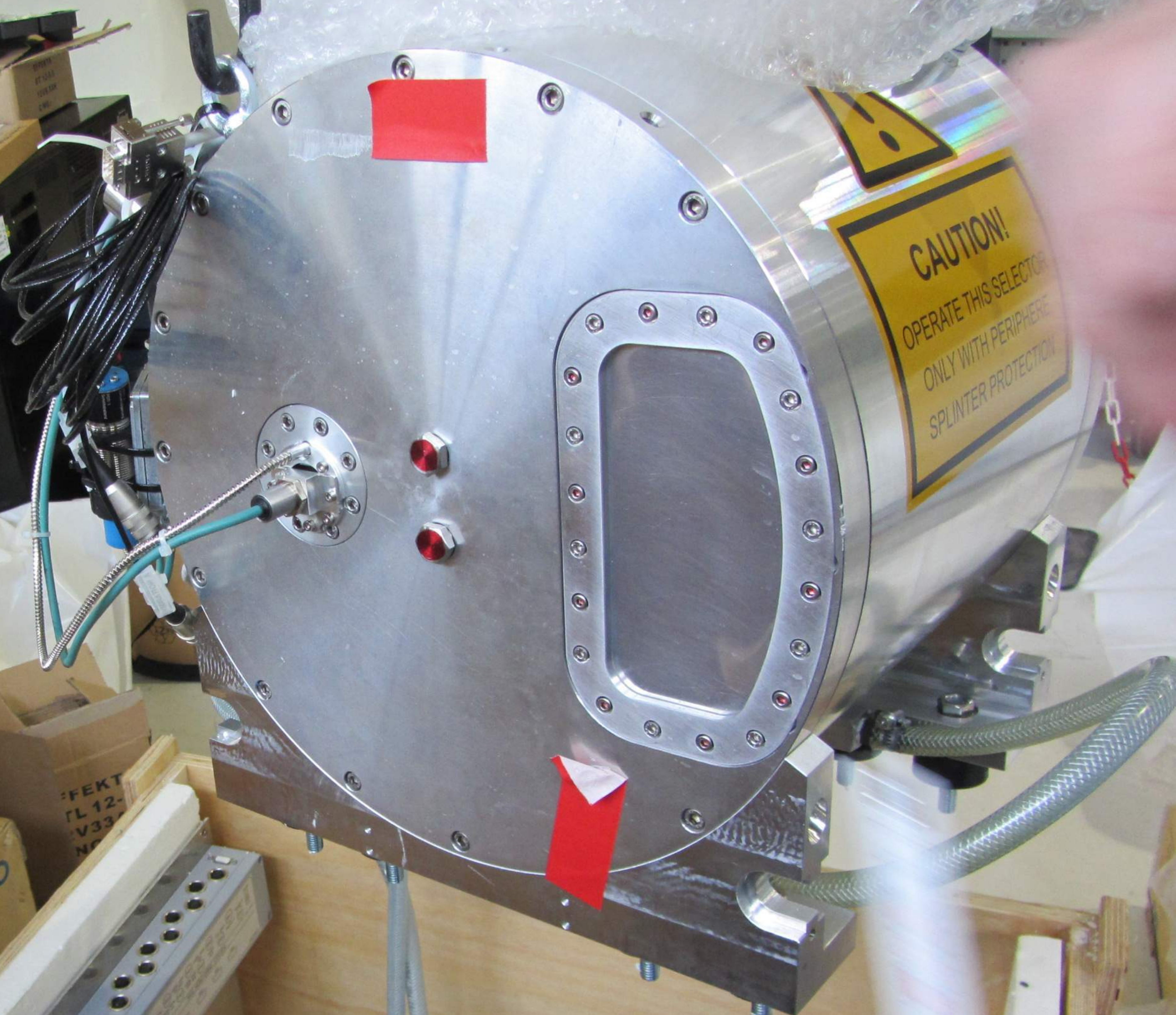}
         \caption{}
         \label{fig:selector2}
     \end{subfigure}
        \caption{Main parts of the velocity selector (VS): (a) the high-speed rotor is equipped with curved lamellae that are made from \textsuperscript{10}B coated carbon fiber. (b) Vacuum housing of the VS. The neutron windows are made from Al.}
\end{figure}

Immediately after the short guide sections, a vertical elevator comprising two positions  is placed. The bottom platform is equipped with a velocity selector (VS) manufactured by Airbus Defence \citep{Airbus}.
The VS is of the helical lamella type \citep{Friedrich1989selector, Wagner1992selector} and its center coincides with the focal point of the horizontally focusing guide system. The rotor is equipped with 50 curved blades with a thickness of 0.4\,mm. They are made from carbon fiber and epoxy with absorbing \textsuperscript{10}B. It is mounted inside a vacuum housing (Figs.\ref{fig:selector1} and \ref{fig:selector2}).

The rotator turns typically with several thousand revolutions per minute (rpm) such that only neutrons with a particular energy band pass the VS. The chosen screw angle of the blades of 23.5$^{\circ}$ results in a minimum wavelength $\lambda$\textsubscript{cut-off}  = 2.35\,\AA{} at the maximum number of revolutions of 28300\,rpm. The wavelength band is $\Delta\lambda/\lambda$ of 30\,\%. The VS effectively suppresses higher harmonics, significantly reduces the background, and increases the signal-to-noise ratio while maintaining a good transmission up to 96\,\% for a non-divergent beam. 
The magnetic guide fields at the rotor that are required to maintain the neutron polarization need to be limited in order to avoid heating by eddy currents.

Thanks to the elevator, the VS can be replaced by a motorized horizontal slit system that is mounted on the upper platform. Even without this slit (i.e. VS in the beam), KOMPASS preserves the concept of having a virtual-source  \citep{Pintschovius1994tas_focusing_mono,Kulda2002in20,Janoschek2010elliptic_guide}
because the distances between the focal point of the parabolic guide and the monochromator and monochromator-sample, respectively, are similar. Inserting the slit at this position perfects the concept as the slit can be adapted to the size of the sample. Measuring with a slit profits from an intensity gain and from an improved beam shape, whereas inserting the velocity selector purifies the spectrum of the incident beam while maintaining essentially the principle of the virtual-source. The mode with the selector driven out will be useful when the higher harmonics are sufficiently suppressed by the curved guide or are not harmful for the experiment.

Next to the elevator, a large flat-coil spin-flipper 
is installed. It is exposed to the vertical magnetic guide field between the elevator and the monochromator. Positioning the flipper in the biological shielding before the monochromator reduces the small but significant influence of the magnetic stray fields stemming from magnetic fields that may be applied at the sample position. A similar flipper can be installed on the optical bench between the sample table and the analyzer.

\subsection{Monochromator and analyzer units} 

\begin{figure}[htb]
     \centering
     \begin{subfigure}[b]{0.55\columnwidth}
         \centering
         \includegraphics[height=3.5cm]{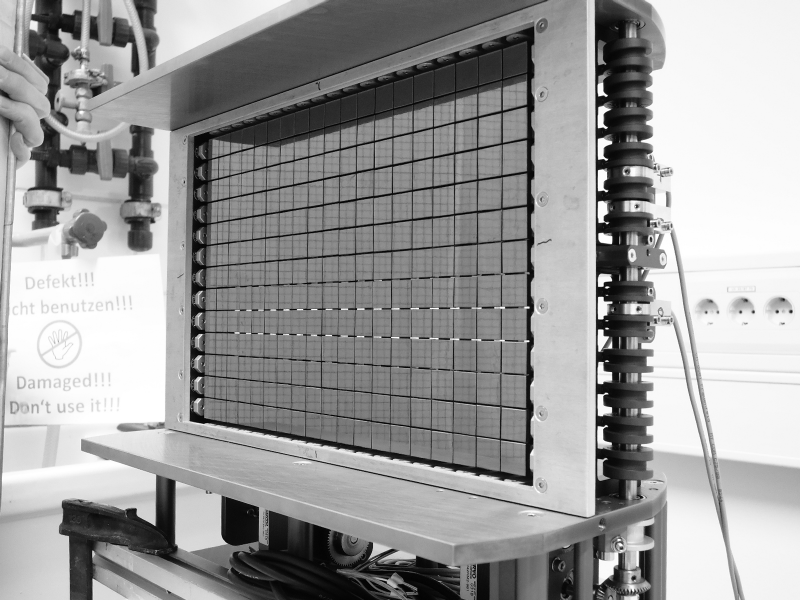}
         \caption{}
         \label{fig:monochromator}
     \end{subfigure}
     \begin{subfigure}[b]{0.4\columnwidth}
         \centering
         \includegraphics[height=3.5cm]{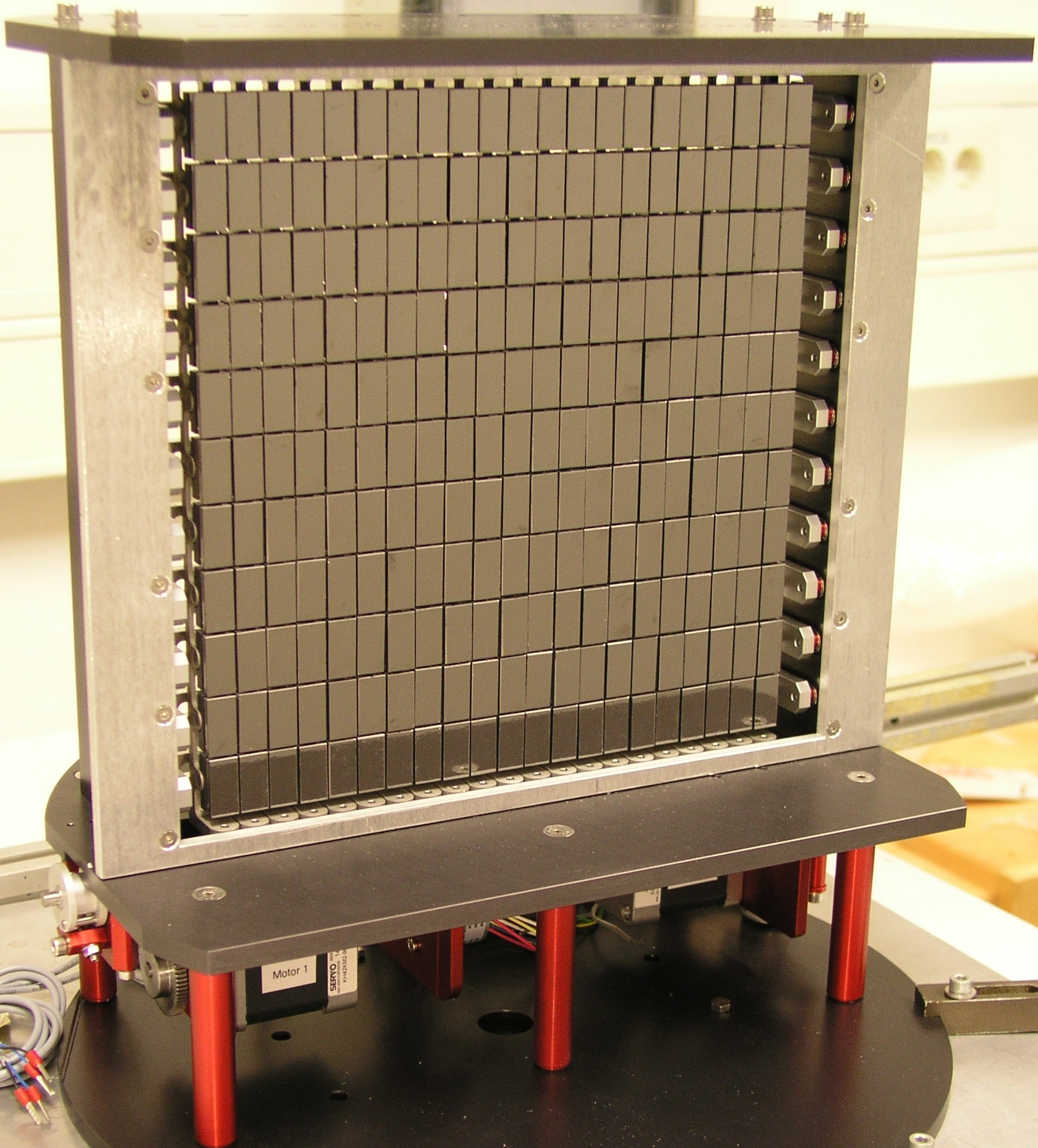}
         \caption{}
         \label{fig:analyser}
     \end{subfigure}
\hfill		 
      \begin{subfigure}[b]{0.4\columnwidth}
         \centering
         \includegraphics[height=3.5cm]{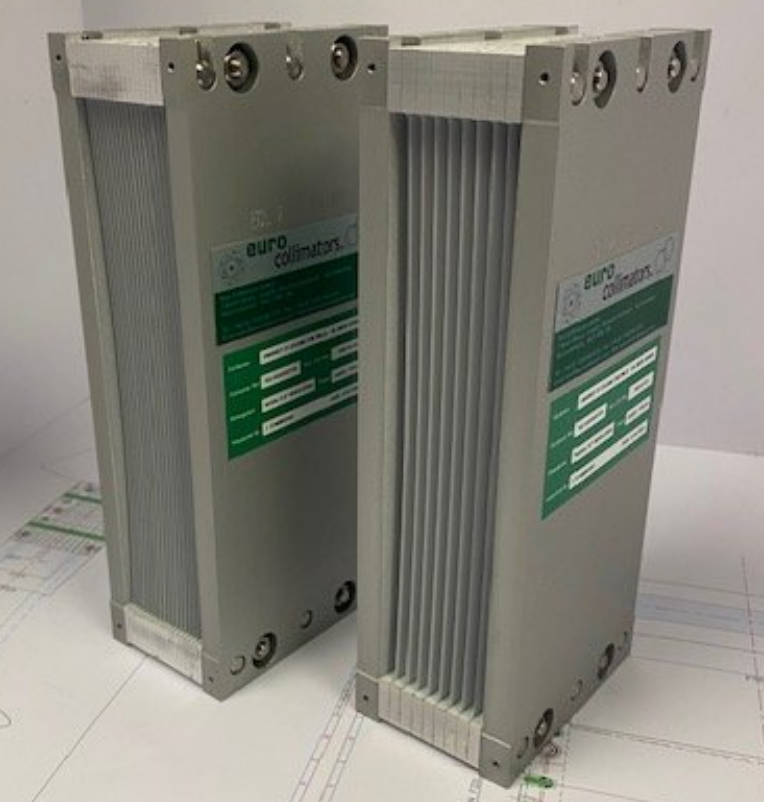}
         \caption{}
         \label{fig:radial_collimators}
     \end{subfigure}
        \caption{Panels (a) and (b) show the doubly-focusing HOPG monochromator and analyzer. (c) Radial collimators before the detector remove the diffuse scattering from the HOPG analyzer.}
\end{figure}

The energy $E_i$ of the incident neutrons 2\,meV $\leq E\textsubscript{in} \leq$ 25\,meV is determined by the take-off angle of the double-focusing monochromator composed of an array of $19 \times 13$ ($w \times h$)  highly oriented pyrolytic graphite (002) crystals. (Fig. \ref{fig:monochromator}). They exhibit a reflectivity close to unity. 
The radii of curvature for horizontal and vertical focusing are remotely adjusted.
The size of the reflecting surface of the monochromator is $275 \times 195$\,mm$^2$ ($w \times h$) sufficient to capture all neutrons from the guide system even at short wavelengths.

The design of the monochromator was carefully optimized to approach a curved surface while limiting the losses arising from the gaps between the HOPG crystals. Detailed McStas simulations performed for different crystal size and mosaic spread predict an increase of the intensity of about 30 \% if one replaces the commonly used crystals with a size of $20 \times 20 \times 2$ \,mm$^3$ ($w \times h \times d$) by crystals with the dimensions $14 \times 14 \times 2.5$ \,mm$^3$ and a mosaic of $0.7^\circ$ \citep{Komarek2011parabolic}. Note that the gap between the crystals in the non-focusing condition is 0.4\,mm.

For background reduction, 1 mm thick platelets consisting of absorbing \textsuperscript{10}B-enriched B\textsubscript{4}C are mounted between the HOPG-crystals and the holders. The vertical guide field across the focusing monochromator is realized by iron yokes that are placed above and below the focusing array. Magnetizing permanent NdFeB-magnets are mounted between the yokes behind the focusing array.

Setting the monochromator-sample distance equal to the monochromator-virtual-source distance and adjusting the width of the virtual source to the sample size yields Rowland focusing conditions if the proper curvature of the monochromator is chosen  \citep{Pintschovius1994tas_focusing_mono, Habicht2012optimizationVS} thus optimizing the performance of the experiment. 

The energy analyzer consists of a similar $220 \times 229$\,mm$^2$ ($w \times h$) double-focusing array of HOPG crystals (Fig. \ref{fig:analyser}). They have a size of $10 \times 20 \times 2$\,mm$^3$ ($w \times h \times d$) with a mosaic of 0.4$^{\circ}$ arranged in 21 columns and 11 rows.  Alternatively, the $170 \times 120$\,mm$^2$ ($w \times h$) fixed-vertically and variably horizontally focusing Heusler analyzer (111) from the instrument PANDA can be installed for polarization analysis \citep{Faulhaber2010lpa}. In this case, the Heusler analyzer replaces or supplements the V-cavity in the secondary spectrometer as discussed below.

Large HOPG analyzers and monochromators naturally produce diffuse scattering close to the specularly reflected beam, thus hampering measurements of weak inelastic signals at small energy transfers \citep{Toft2020HOPG}. The wings of the Bragg peaks of the HOPG crystals contribute to this scattering. This background scattering can be suppressed by inserting Soller collimators, which can be realized at KOMPASS at three positions (Fig. 2(b)). However, they are not adapted to the focusing conditions usually applied to measure weak inelastic scattering. Therefore radial non-oscillating collimators with either 30' or 60' divergence can be placed between the analyzer and the detector (Fig. \ref{fig:radial_collimators}).
The radial collimators use Gd\textsubscript{2}O\textsubscript{3} coated Mylar-foils \citep{Eurocollimators}, which are separated by wedges at the top and bottom and stretched by an aluminum frame. The outer dimensions of the collimators are $65.7 \times 240 \times 100$\,mm ($w \times h \times l$). The number of the absorbing foils is 18 and 8 for the collimator with 30' and 60' divergence, respectively.

\subsection{Biological shielding and beam catcher} 

The main challenge in the construction of the biological shielding at KOMPASS was to obey the strict radiation safety requirements and to maintain a low background below $3 \mu$Sv/h outside the experimental perimeter surrounded by an assembly of shielding walls. The walls are composed of a sandwich of thick borated-epoxy, lead, and borated-polyethylene layers covered with aluminum sheets. These walls are shown in gray in Fig. 1. The door to enter the instrument area is equipped with a interlock safety system that precludes personal presence in the area when the primary shutter is open or when an error scenario occurs.

Especially the intense hard gamma radiation generated by the polarizer and by the selector has to be effectively suppressed despite the very restricted available space and strictly limited floor loading. The design of the shielding was optimized by conducting comprehensive Monte-Carlo simulations prior to the construction \citep{Gruenauer}.
The results revealed that the existing monochromator shielding made from heavy-concrete of the upstream instrument NREX had to be supplemented by a thick lead casing covering the first two polarizing cavities.

The polarizing guide system outside the NREX shielding, the two-position elevator, and the monochromator assembly 
are protected by a biological shielding composed of borated epoxy sheets (50 wt.\% B\textsubscript{4}C) with a thickness of 5 mm that is covered with 240 mm lead and 150 mm borated polyethylene (20 w.\% B\textsubscript{2}O\textsubscript{3}). The latter shielding appears in blue color in Fig. \ref{fig:KOMPASS TAS}. 
Close to the third polarizing cavity, additional 10\,mm thick lead plates are screwed on the outer borated polyethylene plates to suppress the gamma radiation that is generated by photo neutrons from the lead shielding.

The monochromator is shielded by 52 lead blocks that are arranged in two overlapping rows. The blocks are pneumatically lifted to give way for the neutrons, which are reflected from the monochromator with a take-off angle $2\Theta_{\rm mon}$ when varying their energy.
The lead blocks of the inner row are covered with 5\,mm thick sheets of borated aluminum at the side facing the monochromator. The mono-energetic beam passes a rectangular  funnel that is supplemented at the inner side by a thick lead collar, thus reducing the gamma radiation from the monochromator assembly. While changing $2\Theta_{\rm mon}$, the funnel moves together with the diffracted beam, whereas the surrounding lead blocks are sequentially lifted up and down in such a way that only 15 shielding blocks remain lifted at any position. The funnel provides a magnetic guide field and allows a simple mounting of Soller collimators. They are listed in Table. \ref{tab:technical_specification}.

A similar design composed of 36 light pneumatically driven borated polyethylene shielding  blocks is applied to shield the analyzer. Each of the blocks covers an angular range of 9$^{\circ}$. The sides facing the analyzer are covered with sheets of mirrobor (borated rubber containing 82 wt.\% B\textsubscript{4}C). The analyzer shielding requires only one row of shielding blocks, which have an asymmetric zigzag shape preventing the direct view on the analyzer assembly through the shielding.

\begin{figure}[h]
	\centering
	\includegraphics[width=.45\columnwidth]{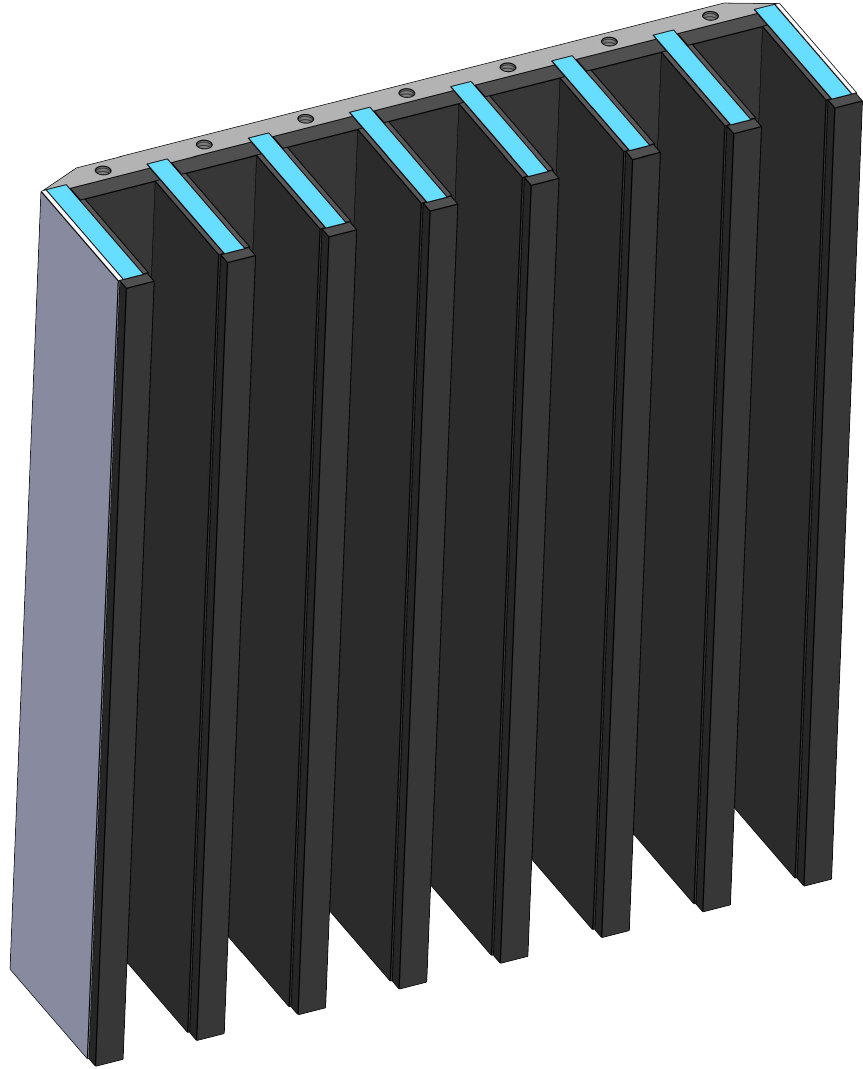}
	\caption{Technical drawing of the direct-beam catcher. Tungsten lamellas (drawn in light blue) are covered with B\textsubscript{4}C sheets shown in black. Gray-colored elements represent aluminium parts of the holding frame. 
	}
	\label{fig:direct_beam_catcher}
\end{figure}

The beam passing the monochromator is absorbed by the direct-beam catcher (see item 11 in Fig. \ref{fig:KOMPASS scheme all}) mounted within the monochromator shielding on the lead wall at the side opposite to the end of the guide system. The catcher consists of a $218 \times 250$\,mm ($w \times h$) aluminium frame with 8 tungsten lamellas of $46 \times 250 \times 6$\,mm ($w \times h \times d$) forming a comb-like structure shown in Fig. \ref{fig:direct_beam_catcher}. The space between the lamellas is covered with 6\,mm thick natural B\textsubscript{4}C sheets, and the lamellas sides and front faces are covered with  1\,mm  and 4\,mm thick B\textsubscript{4}C sheets, respectively. 
The shape of the catcher prevents the backscattering of the direct beam, while the tungsten lamellas and the lead wall behind the catcher shield the gamma radiation generated during the neutron capture. Note that the major part of the direct beam gets absorbed by the 1\,mm thick \textsuperscript{10}B-enriched B\textsubscript{4}C plates mounted behind the HOPG monochromator crystals.

\subsection{Sample table, beamstop and spectral filters} 
\label{sec:Sample_table}

The compact layout of the instrument required designing a multi-axial sample table with small lateral dimensions. It provides many degrees of freedom such as a motorized $(x,y)$\,stage, a set of orthogonal tilting cradles, and a $z$\,stage for the vertical displacement of the sample. In addition, the sample stage serves as a solid mounting base for a versatile sample environment, including a variety of closed-cycle cryostats, two sets of Helmholtz coils, and the {\sl Cryopad} setup for polarization analysis.

The beamstop collecting the part of the monochromatic beam passing through the sample is mounted at the sample table on the side opposite to the monochromator exit (see item 15 in Fig.~\ref{fig:KOMPASS scheme all}). It consists of a 10\,mm thick rectangular B\textsubscript{4}C plate placed on top of a lead plate. For small sample scattering angles $2\Theta_{\rm sample}$, where the optical bench after the sample (item 14 in Fig.~\ref{fig:KOMPASS scheme all}) would collide with the beam stop, it is pneumatically driven down automatically thus allowing to conduct measurements at positive and negative values of $2\Theta_{\rm sample}$ without intervention.

If the VS is not used, higher-order neutrons, background or spurious scattering \cite{Shirane2002book} can be removed by spectral filters that can be mounted on the optical benches. The following filters are available: actively cooled polycrystalline Be and BeO or a pyrolytic graphite higher-order filter.


\subsection{Spin analyzer} 

\begin{figure}[htb]
	\centering
	\begin{subfigure}[]{0.6\columnwidth}
         \centering
         \includegraphics[width=\textwidth]{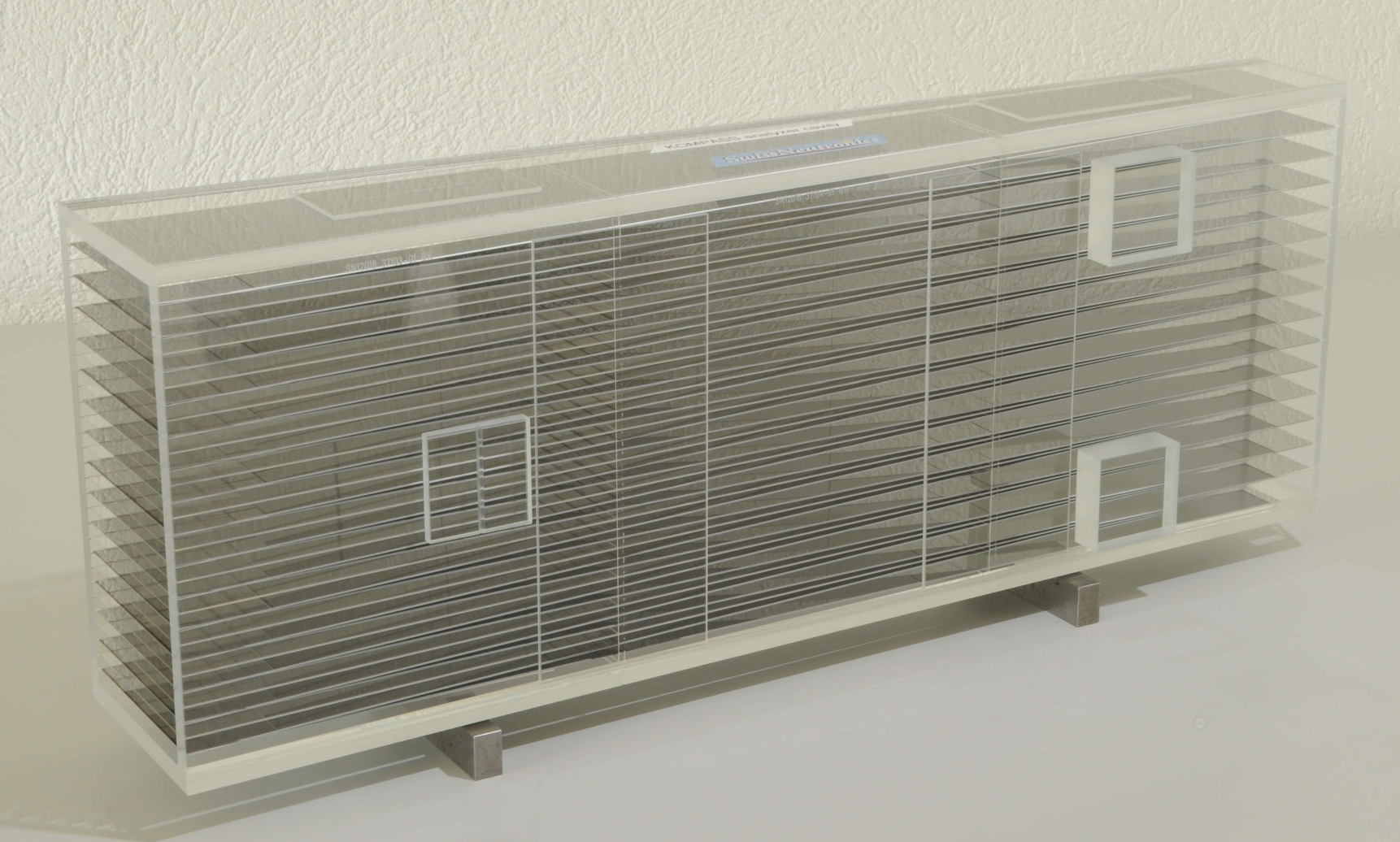}
         \caption{}
         \label{fig:cavity_photo}
     \end{subfigure}
     \begin{subfigure}[]{0.3\columnwidth}
         \centering
         \includegraphics[width=\textwidth]{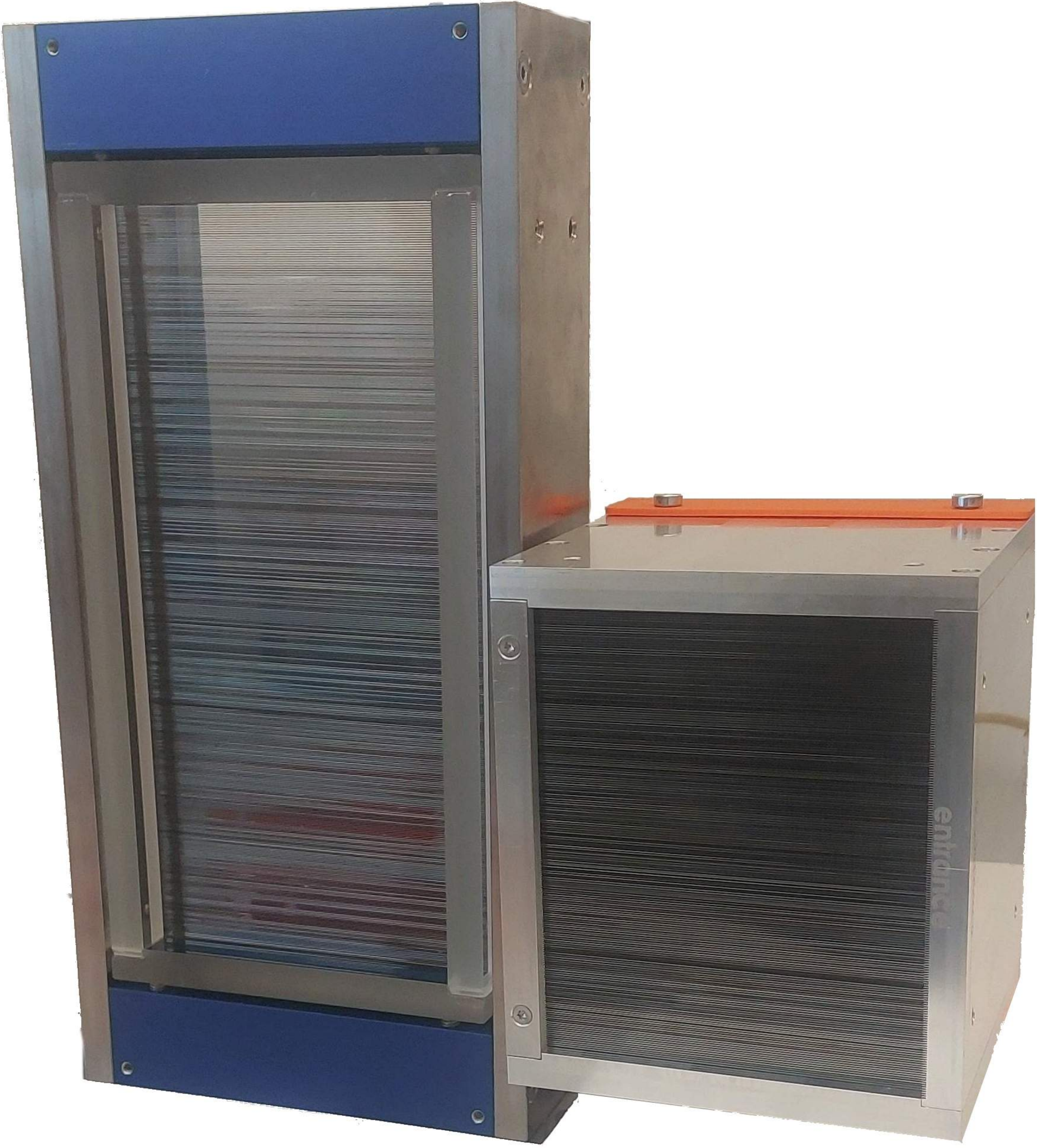}
         \caption{}
         \label{fig:horiz_colim}
     \end{subfigure}          
	\caption{(a) Multichannel polarization analyzer used for the spin analysis of the reflected beam at KOMPASS. The magnetic casing and the surrounding Hallbach array are removed. One can recognize 15 channels, each hosting a SM V-cavity with embedded, double-side Fe/Si-coated SMs on Si substrates. (b) 10’ and 30' solid-state collimators restricting the vertical divergence of the beam can be inserted on front of the polarization analyzer.}  
\end{figure}

The multi-channel V-cavity, shown in Fig. \ref{fig:cavity_photo}, enables the spin analysis of the scattered neutrons at KOMPASS across an area of $100 \times 214$ mm$^2$ ($w \times h$). A single-channel polarizing cavity, as described in section 3.1, would require a long flight path of several meters, which is not available between the analyzer and the detector. Therefore, 15 polarizing V-elements are vertically stacked on top of each other in parallel channels reducing the total length of the analyzer to 650 mm. 


\begin{figure}[h]
	\centering
         \includegraphics[width=.85\columnwidth]{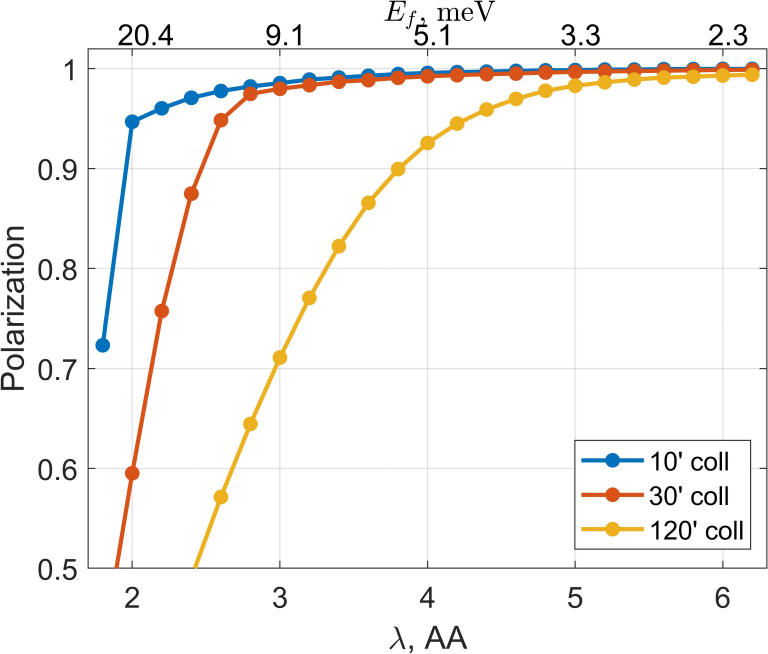}
	\caption{Polarizing efficiency of the polarization analyzer for different vertical divergence of the incoming beam simulated using a divergent source at the sample position. The incoming divergence can be restricted by a collimator in front of the polarization analyzer.}  
	\label{fig:cavity_efficiency}
\end{figure}

Each of the 15 Vs is 602 mm long and covers a beam height of 14 mm. The legs of the Vs consist of 0.3 mm thin silicon wafers coated on both sides with polarizing Fe/Si supermirror $m = 4.2$. The V-elements are separated by horizontally arranged absorbing blades made of uncoated 0.3 mm thin borosilicate glass (Desag 263 T). The trapezoidal absorbing glass body is made of thick, uncoated borofloat glass. It accepts a neutron beam covering an area of $100 \times 214$ mm$^2$ ($w \times h$), with the width tapering to 52 mm towards the exit thus roughly following the horizontal envelope of the focused beam towards the detector (Fig. \ref{fig:cavity_photo}).

According to simulations with McStas, the expected polarization exceeds 96 \% for neutron energies below 15\,meV. Because of the inclination of the legs of the Vs in the vertical direction, the polarization and the transmission of the spin analyzer are largely decoupled from the in-plane $Q$ resolution, which is essentially determined by the beam divergence in the scattering plane given by the horizontal curvature of the analyzer.

However, the polarization efficiency of a V-cavity sensitively depends on the inclination angle between the neutron momentum and the polarizing SMs. Therefore, inserting collimators restricting the vertical divergence of the beam in front of the V-cavity considerably improves the polarization as shown in Fig. \ref{fig:cavity_efficiency}. Two vertical solid-state collimators with 10’ and 30’ divergence shown in Fig.~\ref{fig:horiz_colim} were built to be combined with the SM cavity. The size of the windows amounts to $100 \times 120$\,mm$^2$ ($w \times h$) and $102 \times 215$ \,mm$^2$, for 10' and 30', respectively. The corresponding lengths of the casings are 115\,mm and 60\,mm, for 10' and 30', respectively. 

The collimation is defined by horizontally arranged 0.3\,mm thin silicon wafers that are coated on both sides with absorbing Gd or TiGd layers for 10' and 30', respectively. To reduce the absorption losses, the wafers are separated by 0.3\,mm (and not stacked on top of each other), thus half of the beam propagates through the low-absorbing silicon wafers, while the other half propagates through the air gap between the wafers. Such lamellar solid-state collimators show better transmission when compared with the well-known standard collimators made from Gd\textsubscript{2}O\textsubscript{3}-coated Mylar-foils. In addition, the vanishing waviness of the wafers reduces the beam losses further.

The polarizing SMs are magnetically saturated in a similar way as those of the polarizing guide system. The direction of the guide-field rotates slowly from the vertical direction before the HOPG analyzer to the horizontal direction at the entrance of the analyzer thus rotating the polarization adiabatically. To suppress the intense stray magnetic fields outside the spin analyzer, which would disturb the neighboring neutron-spin-echo spectrometer, the magnetic fields are generated by a self-compensated Halbach-array-like arrangement of permanent magnets and a yoke that was developed in collaboration with Forschungszentrum J{\"u}lich  (Babcock E. et al.~\citep{Babcock202X}). 


The additional magnets generate a total magnetic moment (i.e., product of magnetization times volume) equal to that of the original magnet array, but in the opposite direction. As a result, the magnetic fields of the original and additional magnets cancel each other out at intermediate and longer distances, reducing the stray magnetic fields outside the unit by more than a factor of 100. This reduction is critical for the operation of field-sensitive instruments at adjacent beam ports. In addition to the greatly reduced stray fields outside the analyzer, the magnetic enclosure with the Halbach-like arrangement generates a 25\% enhanced internal magnetic field compared to the simple design, further improving the polarization efficiency of the SMs.

At KOMPASS, it will be possible to combine the spin-analyzing V-cavity in the secondary spectrometer with a polarizing Heusler analyzer, which can be used instead of the usual pyrolytic graphite analyzer described above. Due to the special ratio of nuclear and magnetic Bragg contributions
in the Heusler material, the Heusler polarizer reflects the same neutron polarization with respect to the polarizing magnetic fields at the unit as an SM cavity \citep{Jenni2022}.
Therefore, both polarization systems can easily be combined.

\subsection{Detector options} 
\label{sec:Detector_options}

A commonly used \textsuperscript{3}He tube with 2\,inch diameter and an active length of 120\,mm is the standard detector option at KOMPASS. The detector and spin analyzer are mounted in a large rectangular detector unit equipped with an optical bench on which collimators, apertures, and other beam shaping elements can be mounted. 225\,mm thick sheets of borated polyethylene shield the detector from external radiation. The detector unit (seen as a large blue box on the left in Fig. \ref{fig:KOMPASS TAS}) is placed on the electrically shielded detector electronics box and forms the detector tower unit, which can be attached either directly to the sample stage (in the case of the two-axis mode for diffraction) or to the analyzer tower (for studies with selection of the scattered neutron energy).

Optionally, a 1-inch diameter linear position-sensitive \textsuperscript{3}He detector is available, allowing the neutrons to be recorded as a function of detector height, which can be converted into a vertical divergence and hence vertical $Q$ resolution. Confining the signal to the central part of the detector corresponds to an improved vertical $Q$ resolution, while integrating the whole detector yields more signal at lower $Q$ resolution. Also, the polarization in neutron diffraction can be significantly improved by confining the signal to the central part. The position-sensitive detector allows this selection to be made after the experiment via the analysis software, whereas the vertical height of the detector can normally only be defined by an aperture permanently placed in front of the detector tube.

A set of solid-state collimators with vertically arranged absorbing wafers is used to restrict the horizontal divergence in front of the two detector options. These horizontal collimators improve the $Q$ resolution, however, they will severely cut the intensity in most inelastic studies. In contrast, the losses are significantly smaller in diffraction experiments, where one strives for higher polarization and small background.

To avoid damage during assembly and disassembly of the heavy magnet housing with the sensitive and expensive spin analyzer, a second, shorter detector unit was built, equipped with a detector and an optical bench.  This unit offers the advantage of a shorter analyzer-detector distance of $\sim$1\,m. Within $\sim$20 minutes, the original detector unit with the spin analyzer can be reproducibly replaced by the shorter one, allowing half-polarized experiments or polarized experiments with the Heusler (111) analyzer.

\section{Control of the polarization at the sample}

KOMPASS offers two different setups for controlling the neutron polarization at the sample position. Using a common set of Helmholtz coils, the original vertical magnetic field guiding the neutron polarization after the monochromator is adiabatically rotated in the desired direction at the sample position. The advantage of this solution is its simple technical implementation, although achieving high horizontal fields is not trivial. Since the quantization of the neutron spin is always parallel to the magnetic field, only spin-flip processes can be measured while spin rotation is not accessible. Therefore, this technique is called longitudinal polarization analysis.

\begin{figure}[htb]
	\centering
     \begin{subfigure}[]{0.45\columnwidth}
         \centering
         \includegraphics[width=\textwidth]{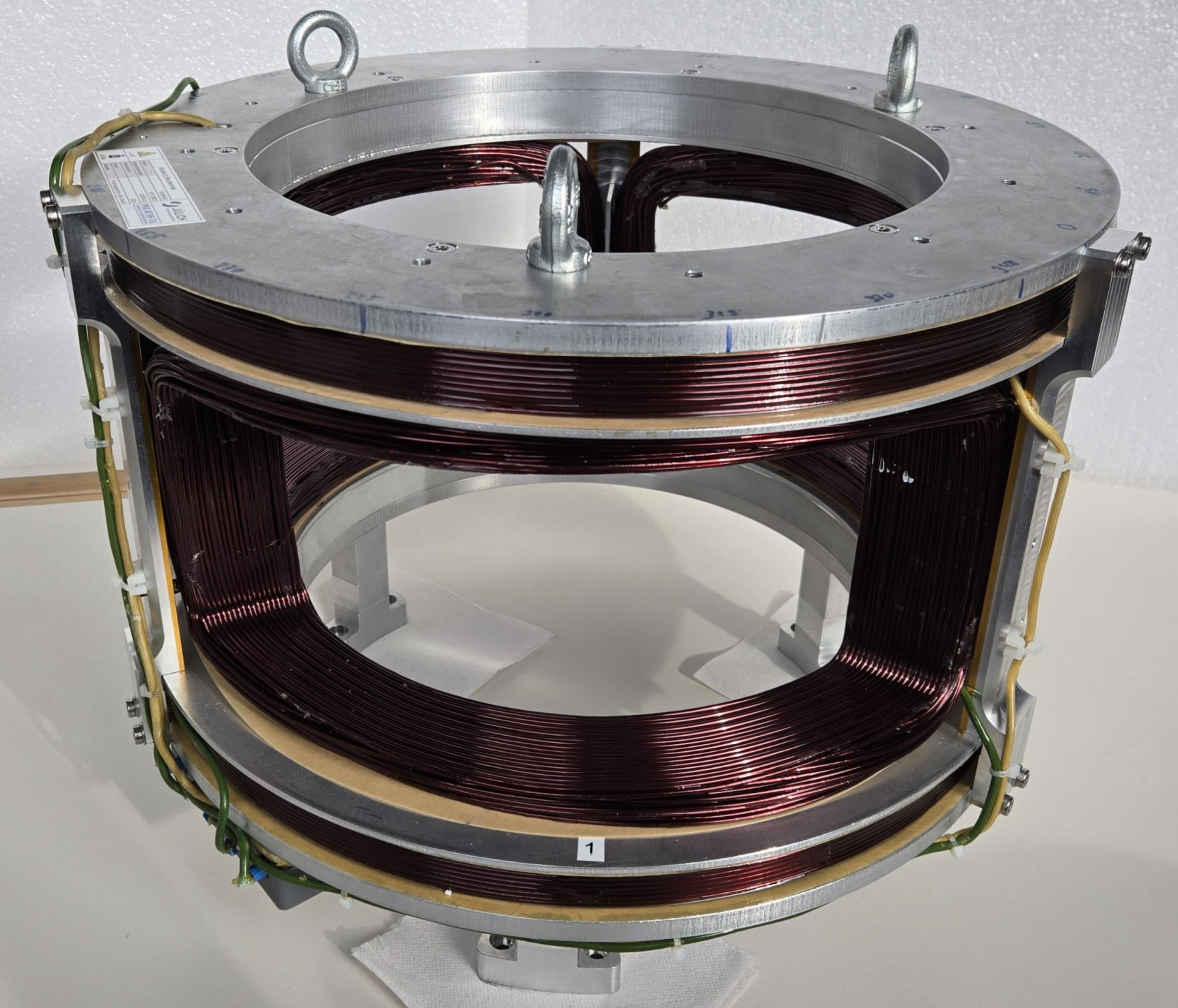}
         \caption{}
     \end{subfigure}
     \begin{subfigure}[]{0.42\columnwidth}
         \centering
         \includegraphics[width=\textwidth]{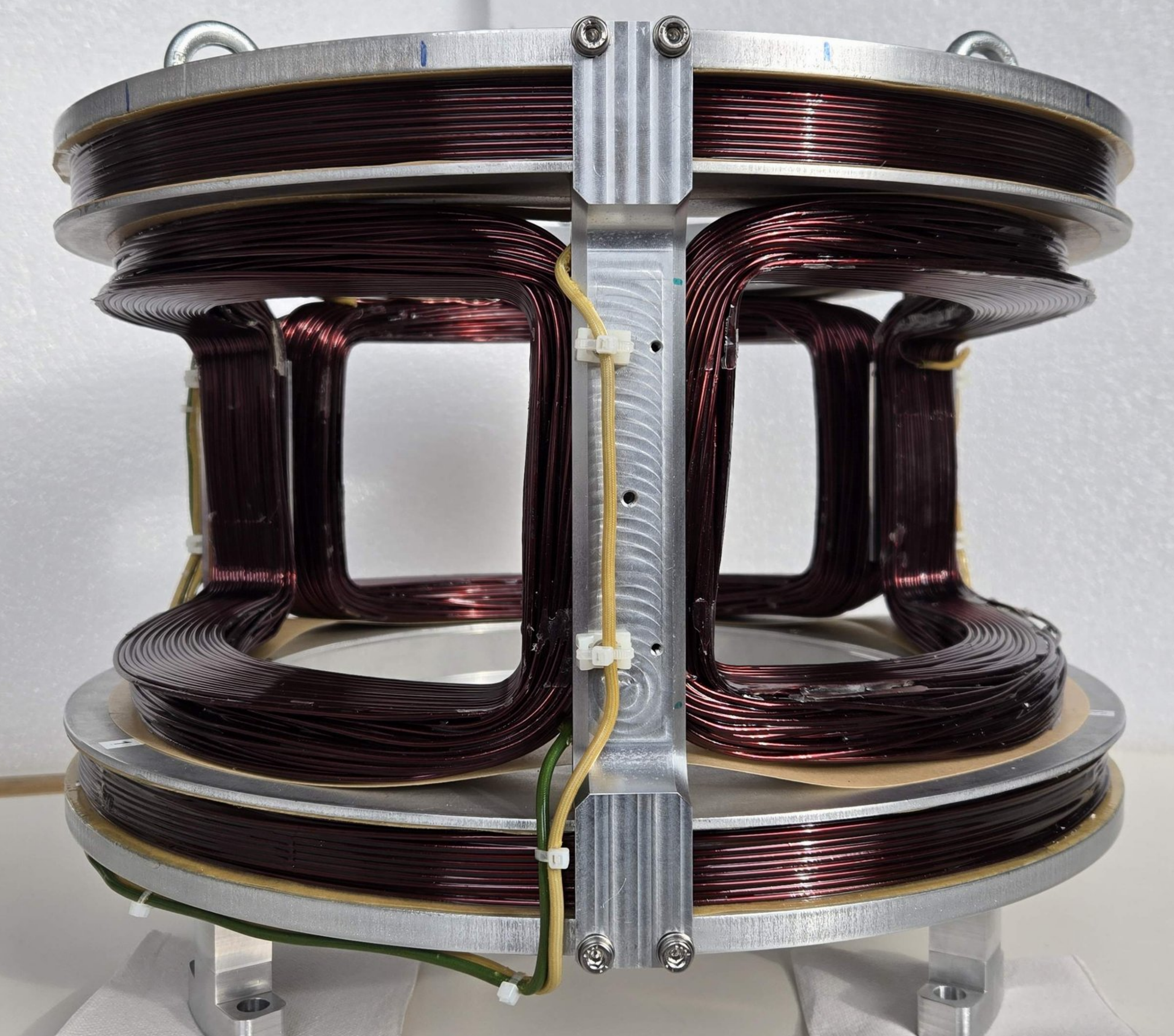}
         \caption{}
     \end{subfigure}        
	\caption{Set of Helmholtz coils allowing one to apply a magnetic guide field of 2\,mT in any direction for longitudinal polarization analysis.} 
	\label{fig:Helmholtz_coils}  
\end{figure}

To generate the magnetic field at the sample, we constructed a compact set of Helmholtz coils that allows to apply a magnetic guide field of 2\,mT in any direction. The system is shown in Fig.~\ref{fig:Helmholtz_coils} and consists of two vertical Helmholtz coils with an outer diameter of 406\,mm and three circularly arranged 120$^\circ$ coil segments with an inner diameter of 285\,mm and vertical opening of 125\,mm. The geometry of the coils was optimized using the COMSOL Multiphysics simulation software \citep{Comsol} to generate a homogeneous field over the typical sample volume up to $40 \times 40 \times 40$\,mm$^3$. In addition, it allows fine alignment of the standard bottom-loader cryostat used at KOMPASS and doesn't require active cooling.

 While it is quite easy to generate sufficiently large vertical guide fields at the sample and along the neutron path, the application of horizontal fields can lead to an unacceptable reduction in the magnetic field along the paths. To avoid such field-free areas, the entire Helmholtz system must be rotated, which is also required to avoid shadowing of the neutron beam by the three columns required for the windings of the three horizontal field coils. Therefore, the Helmholtz system is mounted on a secondary, remotely controlled rotation table of the sample stage. In addition to this conventional system, a setup with a diagonal coil arrangement is also available that does not lead to any shadowing of the beam \citep{Faulhaber2010lpa}.

To overcome the limitations of longitudinal polarization analysis, it is necessary to completely suppress the magnetic field at the sample and to manipulate the polarization of the incoming and outgoing neutrons by guide fields outside the sample space. The {\sl Cryopad} as developed at the Institute Laue-Langevin (ILL) \cite{Lelievre2005,Tasset1999} meets this requirement. Its superconducting  Meissner shield provides an efficient suppression of the magnetic field around the sample. Two nutators combined with two coils for neutron precession in the outer part of the {\sl Cryopad} allow an independent manipulation of the guide fields for the incoming and outgoing neutrons thus adjusting their direction of polarization \cite{Lelievre2005,Tasset1999}. Therefore, one can analyze non-diagonal polarization terms corresponding to the rotation of the neutron polarization thus providing access to transverse polarization terms. This technique is called 3D or spherical polarization analysis.

The third generation {\sl Cryopad} \citep{Lelievre2005,Tasset1999} was acquired in collaboration with the ILL. Only the nutators had to be adapted for KOMPASS. In the nutators, the vertical guide fields are first rotated along the beam and then in any direction perpendicular to the beam when looking towards the sample. For inelastic experiments, it is crucial to guide beams with a large vertical divergence through the nutators, which requires optimization of the large opening between their front magnetic poles. The exit aperture of the nutators at the side pointing to the sample amounts to 40 mm. To define the beam size if necessary, smaller apertures can be inserted between the nutators and the {\sl Cryopad} cryostat. The mechanical mounting of the {\sl Cryopad} on KOMPASS is compatible with small inclinations of the sample goniometers to fine-tune the sample orientation.

\section{Conclusions}

While most TASs provide neutron polarization analysis as an option, the cold-neutron TAS KOMPASS is designed and optimized for the application of polarization techniques, such as longitudinal and spherical polarization analysis, rendering it suitable for studying magnetic correlations in various fields of research. The unique triple supermirror V-cavity permanently incorporated in the parabolically focusing guide system is the heart of KOMPASS delivering a beam with high polarization and good transmission. 

The design of KOMPASS was pursued with the objectives of compactness, advanced focusing, low background, high polarization precision, and flexibility. The reduction of the monochromator-sample-analyzer-detector distances are essential to maintain high statistics and was also required by the little available space.

KOMPASS implements advanced focusing techniques at several places. The large monochromator and analyzer units using highly-oriented pyrolytic graphite are both double focusing with automated control. The guide system is parabolically focusing in its static part while the more decisive end segments can be either straight or focusing to adapt the narrowing of the beam. Focusing guide segments can be optionally inserted directly before and after the sample for studying very small samples.

By transporting only the useful neutrons to the sample, the background can be reduced. This is realized with KOMPASS by the early position of the initial polarizer, by the velocity selector, and by various elements to shape the beam.

KOMPASS offers several modes of operation in its secondary part. First, one may either use the analyzer unit for selection of the final energy or directly mount the detector unit behind the sample tower for diffraction studies. Second, both the elastic and the inelastic configurations can be combined with several polarization options for analyzing the scattered neutrons: none for half-polarized experiments, a SM polarizing cavity, a Heusler unit or the combination of both. Inserting vertical collimators close to the SM polarizing cavity tunes the quality of the polarization analysis, and inserting horizontal collimators at various positions sharpens the $Q$ resolution.

\section{Acknowledgments}
	
The development and installation of KOMPASS is funded by the BMBF through the ErumPro project 05K19PK1. We are grateful to
Peter Link, Thomas Keller, Karin Schmalzl, Wolfgang Schmidt, Eddy Leli{\'e}vre-Berna, Astrid Schneidewind, Paul Steffens, Christian Schanzer and Michael Schneider for stimulating discussions. We also appreciate the electronic workshop of the FZ J{\"u}lich for manufacturing of the Helmholtz coils.

\appendix
\setcounter{figure}{0}                       
\renewcommand\thefigure{A.\arabic{figure}}   

\clearpage

\section{Technical specifications of KOMPASS}

\begin{table*}[H]
\caption{Technical specification of the KOMPASS spectrometer.}
\label{tab:technical_specification}
\begin{tabular*}{.845\textwidth}{@{} |c|c|c|c|c| @{}}
%
\toprule
Incident neutron beam     &\multicolumn{4}{c|}{ 
								\begin{tabular}{@{} c|c|c @{}}
                     2.3 meV < E\textsubscript{n} < 25 meV	
                  &	1.8 < $\lambda$ < 6 \AA{}	
                  &	1.05 < k\textsubscript{i} < 3.5 \AA{}\textsuperscript{-1} \\ 
								 \end{tabular}   
}	\\ 									
\hline

\multicolumn{5}{|c|}{}	\\	
\hline

Polarizer     &\multicolumn{4}{c|}{permanently installed three multichannel V – cavities in series}	\\ 
\hline

\multicolumn{5}{|c|}{}	\\	
\hline

\multicolumn{5}{|c|}{Horizontal focusing of guide sections}	\\
\hline
Static section     &\multicolumn{4}{c|}{parabolic}   \\
\cline{2-5}
Exchangeable long sections     &\multicolumn{4}{c|}{ 
								\begin{tabularx}{10cm}{X|X}
                                     parabolic   &straight     \\ 
								 \end{tabularx}		}	\\ 
\cline{2-5}
Exchangeable short sections    &  \multicolumn{4}{c|}{ 
								\begin{tabularx}{10cm}{X|X|X|X}
                            parabolic   &straight    &20' coll	&40' coll \\ 
								 \end{tabularx}		}	\\ 	
\hline

\multicolumn{5}{|c|}{}	\\	
\hline

Velocity selector     &\multicolumn{4}{c|}{
$\lambda$\textsubscript{cut-off} = 2.35  \AA,  $\Delta\lambda$/$\lambda$ = 30 \%
}	\\ 
\hline

\multicolumn{5}{|c|}{}	\\	
\hline

Monochromator (\textbf{M})	&\multicolumn{4}{c|}{HOPG(002), variable doubly focusing, 275 x 195 mm (w x h)    }   \\
M - crystal array	&\multicolumn{4}{c|}{ 19 x 13 crystals (w x h), each 14 x 14 x 2.5 mm (w x h x d) }   \\
M - scattering angle	    &\multicolumn{4}{c|}{28 $\leq$ 2$\theta_M$ $\leq$ 135 deg}   \\
M-S distance     &\multicolumn{4}{c|}{1.2 m $\leq$ L\textsubscript{MS} $\leq$ 1.67 m}   \\
M - exit collimation     &\multicolumn{4}{c|}{10’, 20’, 40’, 80’ and open}   \\
\hline

\multicolumn{5}{|c|}{}	\\	
\hline

Sample (\textbf{S}) scattering angle     &\multicolumn{4}{c|}{-140 $\leq$ 2$\theta_S$ $\leq$ 140 deg}   \\
S - A distance    &\multicolumn{4}{c|}{1 m $\leq$ L\textsubscript{SA} $\leq$ 1.3 m}   \\
\hline

\multicolumn{5}{|c|}{}	\\	
\hline

\multirow{2}{*}{Analyzer (\textbf{A})}     & \multicolumn{4}{c|}{HOPG(002), variable doubly focusing, 220 x 229 mm (w x h)}\\
\cline{2-5}
{}     & \multicolumn{4}{c|}{Heusler(111), variable horizontal focusing, 170 x 120 mm (w x h)}   \\
\cline{2-5}
\multirow{2}{*}{A - crystal array}     & \multicolumn{4}{c|}{HOPG(002), 11 x 21 crystals (w x h), each 10 x 20 x 2 mm (w x h x d)}\\
\cline{2-5}
{}     & \multicolumn{4}{c|}{Heusler(111), 11 x 3 crystals (w x h), each 17 x 12 x 5 mm (w x h x d)}\\
\cline{2-5}
A - scattering angle     &\multicolumn{4}{c|}{-140 $\leq$ 2$\theta_A$ $\leq$ 140 deg}   \\
\cline{2-5}
\multirow{2}{*}{A - D distance}     & \multicolumn{4}{c|}{with spin-analyzing V – cavity, L\textsubscript{AD} = 1.36 m}\\
{}     & \multicolumn{4}{c|}{half-polarized mode (no cavity), L\textsubscript{AD} = 1 m}   \\
\cline{2-5}
A - entrance collimation     &\multicolumn{4}{c|}{10’, 20’, 40’, 80’ and open}   \\
\hline

\multicolumn{5}{|c|}{}	\\	
\hline

Spin - Analyzer	 &\multicolumn{4}{c|}{Secondary V – cavity, Heusler analyzer or their combination}   \\
\cline{2-5}
{ \begin{tabular}{@{} c @{}} Collimators  in front of \\ secondary V – cavity \end{tabular} }
	 &\multicolumn{4}{c|}{10’ and 30’, restricting vertical divergence}   \\
\hline

\multicolumn{5}{|c|}{}	\\	
\hline

Detector (\textbf{D})     & \multicolumn{4}{c|}{ \o{} 2-inch \textsuperscript{3}He tube or \o{} 1-inch linear position sensitive counter}\\
\cline{2-5}
D - entrance collimation		&\multicolumn{4}{c|}{10’, 20’, 40’, 80’ and open or fixed diaphragms}	\\
\cline{2-5}
Radial collimators	&\multicolumn{4}{c|}{0.5$^{\circ}$ or 1.0$^{\circ}$}	\\
\hline

\multicolumn{5}{|c|}{}	\\	
\hline

Optional filter	&\multicolumn{4}{c|}{Cooled Be and BeO filters, PG filter}	\\
\bottomrule

\end{tabular*}
\end{table*}

\clearpage



\printcredits


\bibliographystyle{elsarticle-num}          

\bibliography{cas-refs_2}

\end{document}